# FEqa: Finite Element Computations on Quantum Annealers

Osama Muhammad Raisuddin, Suvranu De

*Mechanical, Aerospace, and Nuclear Engineering, Rensselaer Polytechnic Institute, 110 8th St. Troy, NY, 12180*

## Abstract:

The solution of physical problems discretized using the finite element methods using quantum computers remains relatively unexplored. Here, we present a unified formulation (FEqa) to solve such problems using quantum annealers. FEqa is a hybrid technique in which the finite element problem is formulated on a classical computer, and the residual is minimized using a quantum annealer. The advantages of FEqa include utilizing a single qubit per degree of freedom, enforcing Dirichlet boundary conditions a priori, reaching arbitrary solution precision, and eliminating the possibility of the annealer generating invalid results. FEqa is scalable on the classical portion of the algorithm due to its Single Program Multiple Data (SPMD) nature and does not rely on ground state solutions from the annealer. The exponentially large number of collocation points used in quantum annealing are investigated for their cosine measures, and new iterative techniques are developed to exploit their properties. The quantum annealer has clear advantages in computational time over simulated annealing, for the example problems presented in this paper solved on the D-Wave machine. The presented work provides a pathway to solving physical problems using quantum annealers.

## 1. Introduction

Quantum computers (QC) show great promise for accelerating scientific computing problems. The properties of superposition, entanglement, and tunneling distinguish QCs from classical computers. The basic unit of computing for a QC, the 'qubit,' can exist in a continuous state of superposition between '0' and '1' until it is measured, whereas a classical 'bit' can be in either of two states '0' or '1'. Two qubits that are entangled are correlated. Two major QC technologies include gate-based quantum computing (Nielsen & Chuang, 2011) and quantum annealing (Finnila et al., 1994). Gate-based quantum computing relies on applying quantum logic gates to manipulate qubits to take advantage of quantum superposition and entanglement properties. Quantum annealing obtains minima by simulating a system of qubits interacting with nearest neighbors to form the Ising Spin Glass Problem (Santoro et al., 2002) and utilizes quantum tunneling to explore low-energy solutions.

Algorithms that exploit the uniqueness of QCs for the solution of linear systems of equations are superior to classical algorithms (Childs et al., 2017; Harrow et al., 2009). While the time complexity of solving an s-sparse system of linear equations with condition number κ on a classical computer is O(Nκ log(1/ε)) using the conjugate gradient method, it is $O(\kappa \log(N)\, poly \log(\frac{1}{\epsilon}))$ using a gate-based quantum computer (Childs et al., 2017), where ε is the accuracy of the approximation. These algorithms rely on noise-free quantum computers. However, existing gate-based Noisy Intermediate Scale Quantum (NISQ) devices used for QC suffer from noise and limited scalability in the number of usable qubits and the number of gates applicable in a sequence.

Gate-based quantum computers are sensitive to noise and require error correction to run fault-intolerant quantum algorithms (Nielsen & Chuang, 2011). The sensitivity of gate-based quantum computers is prohibitive to their application to problems involving large numbers of qubits or algorithms that involve utilizing many quantum gates in succession due to decoherence. On the other hand, Quantum annealers can sample low-energy solutions, though they do not guarantee global minima due to noise and errors. This allows their application towards machine learning, optimization, scientific computing, and quantum chemistry problems (Borle & Lomonaco, 2018; Genin et al., 2019; Li et al., 2018). Hence, in this paper, we use quantum annealers for solving finite element problems.

Quantum annealing is an adiabatic quantum computing method that solves minimization problems faster than classical computers. Quantum annealers simulate an Ising Hamiltonian, which can be programmed to minimize an objective function of interest (Johnson et al., 2011). Quantum annealing has been explored by (Srivastava & Sundararaghavan, 2019) to solve ordinary differential equations using the finite element method. The approach maps each degree of freedom of the discretized system to three qubits. However, the technique is not scalable and may fail due to invalid states being a possible solution of the annealer. Specific choices of the site-interaction and site-site interactions are prescribed for overcoming the issues of these invalid states. However, this approach becomes impractical as the required values of the interaction terms scale unfavorably with the problem size and run into physical constraints on interaction term values imposed by the annealer hardware. The approach does not allow *a priori* enforcement of Dirichlet boundary conditions, further exacerbating the problem of invalid states since enforcement of these boundary conditions also relies on large values of site-interaction terms. The methodology is limited to ordinary differential equations with a symmetric positive definite form, and the imposition of Neumann boundary conditions has not been demonstrated.

Algorithms to solve linear systems of equations using quantum annealers may also be used to solve the linear system of equations obtained from finite element discretization. Such a methodology to solve a linear system of equations using the linear least-squares approach has been developed by (Borle & Lomonaco, 2018). Variables are encoded using the radix-2 approximation for real numbers. This requires $\log_2\left(\frac{1}{\epsilon}\right)$ qubits per degree of freedom to solve to precision $\epsilon$ and exponentially disparate interaction terms to solve the problem with higher precision, limiting practical usage due to physical constraints imposed on the interaction terms by the annealer hardware. This approach also relies on the ground state solution of the system for best results which requires a large number of anneals and is not guaranteed due to noise and discretization errors in annealers (Pearson et al., 2019). The precision of a solution obtained by this approach is limited by the number of qubits used since radix-2 encoding is used.

In this paper, we overcome the limitations of the prior approaches and present a technique (Feqa) to solve finite element problems. Feqa is a hybrid algorithm that formulates the finite element problem as a minimization problem on a classical computer. The minimization is performed iteratively by formulating a series of Ising Hamiltonian problems, of which low energy solutions are sampled from quantum annealers to move in descent directions. Since ground state solutions are not necessary Feqa can utilize NISQ quantum annealers to sample low energy solutions with a low number of anneals per iteration. The precision of solutions obtained using Feqa is not restricted by the precision or range of parameters of the annealer hardware.

Feqa includes the methodology developed by (Borle & Lomonaco, 2018), with the added flexibility over the radix-2 representation by allowing a shift of the zero datum to any arbitrary constant. Previous iterative approaches for ordinary differential equations (Srivastava & Sundararaghavan, 2019) suffer from invalid solutions, including solutions that violate Dirichlet boundary conditions as they use they map the minimization problem directly to the Ising Hamiltonian. We remedy these problems by introducing a modified Ising Hamiltonian along with a mapping procedure that enforces Dirichlet boundary conditions *a priori*. The modified Ising Hamiltonian is equivalent to the Ising Hamiltonian for qubits and allows mapping any general quadratic minimization problem to an Ising Hamiltonian. The modified Ising Hamiltonian is also generalized to qudits (Giorgadze, 2009), higher-dimensional generalizations of qubits, since the grid points formed by qudits provide favorable properties for convergence.

The convergence of gradient-free direct search approaches (Conn et al., 2009) relies on cosine measures. Classical direct search approaches scale poorly in cosine measures for large problems. Cosine measures have not yet been investigated for exponentially large grid points realizable in quantum annealing. We show that the cosine measure of the grid points formed by qubits ($2^N$ Search) does not provide any advantage over classical pattern search in cosine measure, but the grid points formed by qudits yield an exponential improvement in cosine measure. However, since annealer hardware is limited to qubits, we present nested iterations to realize the gridpoints of higher-order qudits using qubits. We demonstrate that nested iterations are resource-intensive in both qubits and qubit connectivity. To remedy this problem, we present hyperoctant search as a heuristic technique that uses additional anneals to improve the search results obtained using $2^N$ Search.

We benchmark quantum annealing against simulated annealing using time-to-target metrics developed by (King et al., 2015). These metrics are suited for the application since they allow benchmarking of algorithms that do not rely on ground state solutions of the Ising Hamiltonian.

The paper is organized as follows. Section 2 provides the mathematical formulation for the finite element problem. Section 3 introduces quantum annealing and the Ising Hamiltonian Problem and modifications to generalize the Ising Hamiltonian. We also provide the methodology to map a finite element problem to an Ising Hamiltonian problem. Section 4 introduces cosine measures for the exponential spaces realizable using quantum annealing, iterative procedures to solve finite element problems using quantum annealers, and the time-to-target metric. Section 5 presents results obtained using quantum annealing and compares them with simulated annealing results. Discussion and future directions are provided in Section 6.

## 2. Finite element problem

We consider the following linear finite element problem defined on the open bounded domain $\Omega_h \in \mathbb{R}^d, d \in \{1,2,3\}$ with boundary $\Gamma_h$

Find $\boldsymbol{u} \in \mathbb{R}^N$ such that

$$\boldsymbol{\Psi}(\boldsymbol{u}) = \mathbf{A}\boldsymbol{u} - \mathbf{b} = \mathbf{0} \tag{1}$$

subject to $\boldsymbol{u} = \boldsymbol{u}_g$ on $\Gamma_g$

where $\boldsymbol{u}$ is a vector of nodal unknowns, $\boldsymbol{A} \in \mathbb{R}^{N \times N}$ is the system matrix, $\boldsymbol{b} \in \mathbb{R}^N$ is the forcing function and $N$ is the number of degrees of freedom of the discretized problem and $\Gamma_g$ is the Dirichlet boundary.

For a symmetric, positive definite $A$,

$$\boldsymbol{u} = \underset{\boldsymbol{v} \in \mathbb{R}^N}{\text{argmin}}\, \mathcal{F}(\boldsymbol{v}) \tag{2}$$

with the following functional

$$\mathcal{F}(\boldsymbol{v}) = \frac{1}{2}\boldsymbol{v}^T \boldsymbol{A} \boldsymbol{v} - \boldsymbol{v}^T \boldsymbol{b} \tag{3}$$

For the general case of nonsymmetric $A$,

$$\boldsymbol{u} = \underset{\boldsymbol{v} \in \mathbb{R}^N}{\text{argmin}}\, \mathcal{G}(\boldsymbol{v}) \tag{4}$$

with

$$\mathcal{G}(\boldsymbol{v}) = \|\Psi(\boldsymbol{v})\|_2 \tag{5}$$

where $\|\cdot\|_2$ is the L₂ norm.

## 3. Quantum annealing for finite element problems

First, we will provide definitions of the fundamental computational units of quantum computing:

**Definition 1:** A qubit or quantum bit $\overset{2}{q}_j$ is a system whose quantum state $\overset{2}{q}_j \in \mathcal{H}_j$ where $\mathcal{H}_j \cong \mathbb{C}^2$ can be described by a superposition of two orthogonal eigenstates labeled as $|-1\rangle$ and $|1\rangle$, and can classically take on 2 states $\overset{2}{q}_j \in \mathbb{R}$ described as eigenstates $\overset{2}{q}_j \in \{1, -1\}$.

**Note**: The labels $|1\rangle, |-1\rangle$ correspond to physical properties, e.g., spin-up or spin-down, $|\uparrow\rangle, |\downarrow\rangle$ which depend on the choice of arbitrary measurement axes.

**Definition 2**: A register of $N$ qubits $\overset{2}{\boldsymbol{q}}$ has a quantum state $\overset{2}{\boldsymbol{q}} \in \mathcal{H}$ where $\mathcal{H} \cong \mathcal{H}_1 \otimes \mathcal{H}_2 \ldots \mathcal{H}_{N-1} \otimes \mathcal{H}_N \cong \mathbb{C}^{2^N}$, and can classically take on $2^N$ states $\overset{2}{\boldsymbol{q}} \in \mathbb{R}^N$ described as eigenstates of the individual qubits.

**Definition 3:** A qudit $\overset{d}{q}_k$ is a system whose quantum state $\overset{d}{q}_k \in \mathcal{H}_k$ where $\mathcal{H}_k \cong \mathbb{C}^d$ can be described by a superposition of $d > 2$ orthogonal eigenstates labeled as the integral stepped values

$|-m\rangle, |-m+1\rangle, \ldots, |m-1\rangle, |m\rangle$, and can classically take on d states $\overset{d}{q}_k \in \mathbb{R}$ described as eigenstates $\{-m, -m+1, \ldots m-1, m\}$.

Qudits extend the idea of quantum computing to higher dimensions (Wang et al., 2020) and have been experimentally realized (O'Sullivan-Hale et al., 2005).

**Note**: A qutrit is denoted as $\overset{3}{q}$ and a ququart as $\overset{4}{q}$.

**Definition 4:** A register of $N$ qudits $\overset{d}{\boldsymbol{q}}$ has a quantum state $\overset{d}{\boldsymbol{q}} \in \mathcal{H}$ where $\mathcal{H} \cong \mathcal{H}_1 \otimes \mathcal{H}_2 \ldots \mathcal{H}_{N-1} \otimes \mathcal{H}_N \cong \mathbb{C}^{d^N}$ and can classically take on $d^N$ states $\overset{d}{\boldsymbol{q}} \in \mathbb{R}^N$ described as eignestates of the individual qudits.

Quantum annealers are purpose-built devices that ideally provide the ground state solution ($\overset{\widetilde{2}}{\boldsymbol{q}}$) for the programmed Ising Hamiltonian problem (Brooke et al., 1999):

$$\widetilde{\overset{2}{q}} = \arg\min_{\overset{2}{q}\in\mathbb{R}^N} \tilde{E}\left(\overset{2}{q}\right) \tag{6}$$

where the Ising Hamiltonian $\tilde{E}\left(\overset{2}{q}\right): \mathbb{R}^N \to \mathbb{R}$ for a system of $N$ qubits is defined as

$$\tilde{E}\left(\overset{2}{q}\right) = \overset{2}{q}^T \tilde{J} \overset{2}{q} + \overset{2}{q}^T \tilde{h} \tag{7}$$

where $\tilde{J} \in \mathbb{R}^{N\times N}$ is strictly upper diagonal and $\tilde{h} \in \mathbb{R}^N$. $\tilde{h}_i$ are qubit biases, also referred to as site interactions, and $\tilde{J}_{i,j}$ are coupling strengths between qubits, also referred to as site-site interactions. $\tilde{h}$ and $\tilde{J}$ are the programmable parameters of the Ising Hamiltonian problem, and the classical state $\widetilde{\overset{2}{q}}$ is the output. Although $\tilde{h}_i$ and $\tilde{J}_{i,j} \in (-\infty, \infty)$, they are linearly scaled to fit the programmable range of specific annealer hardware due to hardware constraints.

**Note:** In practice, the global minimum of the Ising Hamiltonian may not be obtained by the quantum annealer due to noise and discretization errors (Pearson et al., 2019). As with any quantum system, the outcome is a probability distribution over the possible outcomes, and quantum annealing produces low eigenenergy eigenstates of the Ising Hamiltonian with high probability. Current annealers can provide low eigenenergy eigenstates if not the ground state (King et al., 2015).

We will compare the eigenstates and eigenenergies obtained using quantum annealing, in which eigenstates are obtained in a single step on the annealer hardware, with eigenstates and eigenenergies obtained using simulated annealing, in which eigenstates are obtained iteratively in $k$ steps, also referred to as sweeps, as shown in Algorithm 1. For the sake of completeness, the algorithm for simulated annealing is summarized in Algorithm 1 below ((Laarhoven Aarts, E. H. L., 1987).

Algorithm 1. Pseudo-code of simulated annealing

1. **Input $q_0, k_{max}$, Initialization:** $k = 0, q = q_0$
2. **while** $k < k_{max}$
3.    $T = temperature(1 - (k+1)/k_{max})$
4.    $q_{new} = neighbor(q)$
5.    if $P(E(q), E(q_{new}), T) \geq random(0,1)$
6.       $q = q_{new}$
7.    k = k+1
8. **end while**
9. **Output:** $q$

### 3.1. Modified Ising Hamiltonian

We define the following modified Ising Hamiltonian problem as

$$\widetilde{\overset{d}{q}} = \arg\min_{\overset{d}{q}} E\left(\overset{d}{q}\right) \tag{8}$$

with the modified Ising Hamiltonian $E\left(\overset{d}{q}\right): \mathbb{R}^N \to \mathbb{R}$ for both qubits and qudits defined as:

$$E\left(\overset{d}{q}\right) = \overset{d}{q}^T J \overset{d}{q} + \overset{d}{q}^T h + \overset{d}{q}^T S \overset{d}{q} + c \tag{9}$$

where $J \in \mathbb{R}^{N \times N}$ is hollow, $h \in \mathbb{R}^N$, $S \in \mathbb{R}^{N \times N}$ is diagonal, and $c \in \mathbb{R}$, and $d \geq 2$.

The modified Ising Hamiltonian maps $d^N$ eigenstates of $\overset{d}{q}$ to $d^N$ eigenenergies. It contains all the terms in a general quadratic functional including a constant, which allows any arbitrary quadratic functional to be mapped directly to the modified Ising Hamiltonian for an equivalent minimization problem over $d^N$ eigenstates.

**Note:** $\arg\min \tilde{E}\left(\overset{2}{q}\right) = \arg\min E\left(\overset{2}{q}\right)$ since $\overset{2}{q}_i \in \{+1, -1\}$ and $\overset{2}{q}^T S \overset{2}{q} = trace(S)$ is constant and independent of $\overset{2}{q}$, both $\arg\min \tilde{E}\left(\overset{2}{q}\right)$ and $\arg\min E\left(\overset{d}{q}\right)$ are independent of constant terms, and $\tilde{J} = upper(J) + lower(J)^T$. These additional terms and relaxation of the restrictions on $\tilde{J}$ allows a direct mapping of the discretized finite element problem in Equations (3) and (5) to the modified Ising Hamiltonian. $S$ and $c$ can simply be discarded since the minimization problem is independent of constants.

### 3.2. Mapping a Quadratic Functional to the Ising Hamiltonian

Unconstrained minimization of general quadratic functionals like $\mathcal{F}(u)$ and $G(u)$ can be mapped to either an equivalent Ising Hamiltonian problem, hereafter referred to as *direct annealing*, or an iterative process that uses Ising Hamiltonian problems to iteratively converge to $u$, hereafter referred to as *iterative annealing*.

Direct annealing approaches encode each element of $u$ as a set of qubits, e.g., the radix-2 approximation, map the quadratic functional to an Ising Hamiltonian problem and attempt to find the ground state solution for $u_{approx.}$ (Borle & Lomonaco, 2018). This approach requires only one Ising Hamiltonian to be formulated and minimized. However, the larger solution space arising from several qubits assigned to each unknown and the requirements to reach the ground state requires a significant number of anneals, the annealer graphs produced using this approach are typically dense, and the components of the parameters $\tilde{h}$ and $\tilde{J}$ have exponentially disparate values which cannot be implemented on real hardware. Furthermore, precision is limited to the number of qubits assigned to each unknown.

Iterative annealing approaches use an iterative procedure to get better estimates of $u$ by solving a series of Ising Hamiltonian problems until a convergence criterion is met (Chang et al., 2019; Srivastava & Sundararaghavan, 2019). It is a hybrid approach that can use a single qubit per unknown and produces sparser graphs than those for direct annealing for the same problem. Ising Hamiltonian problems are formulated using a classical computer and minimized using a quantum annealer. The annealer results are processed on a classical computer for the next iteration. The ground state solution is not necessary for convergence; thus, fewer anneals are needed.

We propose a general iterative minimization technique of a quadratic functional $\mathcal{H}(u): \mathbb{R}^N \to \mathbb{R}$ using the following iteration

$$u_{i+1} = u_i + \Delta_i + D_i \overset{d}{q}_i \tag{10}$$

where $i$ is the iteration number, $\boldsymbol{D}_i \in \mathbb{R}^{N \times N}$ is a matrix used to control the size and shape of the search neighborhood around $\boldsymbol{u}_i \in \mathbb{R}^N$, and $\boldsymbol{\Delta}_i \in \mathbb{R}^N$ is a vector to translate the search neighborhood. Each unknown $u_{i_k}$ is mapped to a single qudit $\overset{d}{q}_{i_k}$.

For functionals of the form defined in Equation (3), the substitution Equation (10) leads to

$$\mathcal{F}(\boldsymbol{u}_{i+1}) = \tfrac{1}{2} \overset{d}{\boldsymbol{q}}_i^T \boldsymbol{D}_i^T A \boldsymbol{D}_i \overset{d}{\boldsymbol{q}}_i + \overset{d}{\boldsymbol{q}}_i^T \left( \boldsymbol{D}_i^T (A(\boldsymbol{u}_i + \boldsymbol{\Delta}_i) - \boldsymbol{b}) \right) + (\boldsymbol{u}_i + \boldsymbol{\Delta}_i)^T \left( \tfrac{1}{2} A(\boldsymbol{u}_i + \boldsymbol{\Delta}_i) - \boldsymbol{b} \right) \quad (11)$$

Equation (11) has the form of Equation (8) and can be minimized over $d^N$ eigenstates of $\overset{d}{\boldsymbol{q}}$ using the modified Ising Hamiltonian problem by sampling for either the ground state or low energy eigenstates. By comparing with the terms in Equation (8) we obtain

$$\boldsymbol{J}_i = \tfrac{1}{2} \boldsymbol{D}_i^T A \boldsymbol{D}_i - Diag\left(\tfrac{1}{2} \boldsymbol{D}_i^T A \boldsymbol{D}_i\right) \quad (12)$$

$$\boldsymbol{h}_i = \boldsymbol{D}_i^T (A(\boldsymbol{u}_i + \boldsymbol{\Delta}_i) - \boldsymbol{b}) \quad (13)$$

$$\boldsymbol{S}_i = Diag\left(\tfrac{1}{2} \boldsymbol{D}_i^T A \boldsymbol{D}_i\right) \quad (14)$$

$$c_i = (\boldsymbol{u}_i + \boldsymbol{\Delta}_i)^T \left(\tfrac{1}{2} A(\boldsymbol{u}_i + \boldsymbol{\Delta}_i) - \boldsymbol{b}\right) \quad (15)$$

Similarly, for functionals of the form defined by Equation (4), the substitution yields

$$\mathcal{G}(\boldsymbol{u}_{i+1}) = \overset{d}{\boldsymbol{q}}_i^T \boldsymbol{D}_i^T A^T A \boldsymbol{D}_i \overset{d}{\boldsymbol{q}}_i + 2\overset{d}{\boldsymbol{q}}_i^T \boldsymbol{D}_i^T A^T (A(\boldsymbol{u}_i + \boldsymbol{\Delta}_i) - \boldsymbol{b}) + (\boldsymbol{u}_i + \boldsymbol{\Delta}_i)^T A^T A (\boldsymbol{u}_i + \boldsymbol{\Delta}_i) - 2(\boldsymbol{u}_i + \boldsymbol{\Delta}_i)^T A^T \boldsymbol{b} + \boldsymbol{b}^T \boldsymbol{b} \quad (16)$$

and comparing with Equation (8) yields

$$\boldsymbol{J}_i = \boldsymbol{D}_i^T A^T A \boldsymbol{D}_i - Diag(\boldsymbol{D}_i^T A^T A \boldsymbol{D}_i) \quad (17)$$

$$\boldsymbol{h}_i = 2 \boldsymbol{D}_i^T A^T (A(\boldsymbol{u}_i + \boldsymbol{\Delta}_i) - \boldsymbol{b}) \quad (18)$$

$$\boldsymbol{S}_i = Diag(\boldsymbol{D}_i^T A^T A \boldsymbol{D}_i) \quad (19)$$

$$c_i = (\boldsymbol{u}_i + \boldsymbol{\Delta}_i)^T A^T A(\boldsymbol{u}_i + \boldsymbol{\Delta}_i) - 2(\boldsymbol{u}_i + \boldsymbol{\Delta}_i)^T A^T \boldsymbol{b} + \boldsymbol{b}^T \boldsymbol{b} \quad (20)$$

In this paper, we make the following choice for $\boldsymbol{D}_i$

$$\boldsymbol{D}_i = \alpha_i (\boldsymbol{I} - \boldsymbol{G}) \quad (21)$$

where $\alpha_i$ is a scaling constant controlling the search neighborhood, described in more detail in Section 4, and $\boldsymbol{G}$ is the following diagonal matrix

$$G_{ii} = \begin{cases} 1, & u_i \in \Gamma_g \\ 0, & u_i \notin \Gamma_g \end{cases} \quad (22)$$

This choice of $\boldsymbol{D}$ collapses the search dimensions associated with Dirichlet boundary conditions to enforce them *a priori*. The interactions $h_g$, $J_{ig}$, $J_{gi}$, and $S_{gg}$ for the qudits $\overset{d}{q}_g$ associated with $\boldsymbol{u}_g$ are zero and therefore do not appear in the minimization problem. This allows the qudits $\overset{d}{q}_g$ to be dropped from the Ising Hamiltonian Problem, leading to a smaller and less dense annealer graph.

### 3.3. Nested Iterations

Equation (11) can be rewritten as

$$\mathcal{F}(u_{i+1}) = \frac{1}{2}\overset{d}{q_i}^T D_i^T A D_i \overset{d}{q_i} + \overset{d}{q_i}^T \left(D_i^T(A(u_i + \Delta_i) - b)\right) + \mathcal{F}(u_i) + \frac{1}{2}u_i^T A(\Delta_i) + (\Delta_i)^T \left(\frac{1}{2}A(u_i + \Delta_i) - b\right) \quad (23)$$

$$\mathcal{F}(u_{i+2}) = \frac{1}{2}\overset{d}{q}_{i+1}^T D_{i+1}^T A D_{i+1} \overset{d}{q}_{i+1} + \overset{d}{q}_{i+1}^T \left[D_{i+1}^T(A(u_{i+1} + \Delta_{i+1}) - b)\right] + \mathcal{F}(u_{i+1}) + \frac{1}{2}u_{i+1}^T A(\Delta_{i+1}) + (\Delta_{i+1})^T \left(\frac{1}{2}A(u_{i+1} + \Delta_{i+1}) - b\right) + \frac{1}{2}(u_i + \Delta_i)^T A(\Delta_{i+1}) + (\Delta_{i+1})^T \left(\frac{1}{2}A(u_{i+1} + \Delta_{i+1}) - b\right) \quad (24)$$

Leading to a nested iteration of the form

$$\mathcal{F}(u_{i+2}) = \begin{bmatrix} \overset{d}{q_i} \\ \overset{d}{q}_{i+1} \end{bmatrix}^T \begin{bmatrix} \frac{1}{2}D_i^T A D_i & 0 \\ D_{i+1}^T A D_i & \frac{1}{2}D_{i+1}^T A D_{i+1} \end{bmatrix} \begin{bmatrix} \overset{d}{q_i} \\ \overset{d}{q}_{i+1} \end{bmatrix} + \begin{bmatrix} \overset{d}{q_i} \\ \overset{d}{q}_{i+1} \end{bmatrix}^T \begin{bmatrix} D_i^T(A(u_i + \Delta_i) - b) + \frac{1}{2}D_i^T A \Delta_{i+1} \\ D_{i+1}^T(A(u_i + \Delta_i + \Delta_{i+1}) - b) \end{bmatrix} + \mathcal{F}(u_i) + \frac{1}{2}u_i^T A(\Delta_i) + (\Delta_i)^T \left[\frac{1}{2}A(u_i + \Delta_i) - b\right] + \frac{1}{2}(u_i + \Delta_i)^T A(\Delta_{i+1}) + (\Delta_{i+1})^T \left[\frac{1}{2}A(u_{i+1} + \Delta_{i+1}) - b\right] \quad (25)$$

This can be minimized in one step using the modified Ising Hamiltonian with

$$J_{(i,i+1)} = \begin{bmatrix} \frac{1}{2}D_i^T A D_i & 0 \\ D_{i+1}^T A D_i & \frac{1}{2}D_{i+1}^T A D_{i+1} \end{bmatrix} - trace\left(\begin{bmatrix} \frac{1}{2}D_i^T A D_i & 0 \\ D_{i+1}^T A D_i & \frac{1}{2}D_{i+1}^T A D_{i+1} \end{bmatrix}\right) \quad (26)$$

$$h_{(i,i+1)} = \begin{bmatrix} D_i^T(A(u_i + \Delta_i) - b) + \frac{1}{2}D_i^T A \Delta_{i+1} \\ D_{i+1}^T(A(u_i + \Delta_i + \Delta_{i+1}) - b) \end{bmatrix} \quad (27)$$

$$S_{(i,i+1)} = trace\left(\begin{bmatrix} \frac{1}{2}D_i^T A D_i & 0 \\ D_{i+1}^T A D_i & \frac{1}{2}D_{i+1}^T A D_{i+1} \end{bmatrix}\right) \quad (28)$$

$$c = \mathcal{F}(u_i) + \frac{1}{2}u_i^T A(\Delta_i) + (\Delta_i)^T \left[\frac{1}{2}A(u_i + \Delta_i) - b\right] + \frac{1}{2}(u_i + \Delta_i)^T A(\Delta_{i+1}) + (\Delta_{i+1})^T \left[\frac{1}{2}A(u_{i+1} + \Delta_{i+1}) - b\right] \quad (29)$$

This procedure can be used to obtain the $n^{th}$ nested iteration formulae for $\mathcal{F}(u_{i+n})$ and $\mathcal{G}(u_{i+n})$. The general forms for a choice of $\Delta_i, \Delta_{i+1}, \ldots, \Delta_{i+n-1} = 0$ are

$$\mathcal{F}(u_{i+n}) = \sum_{j=0}^{n-1} \frac{1}{2}\overset{d}{q}_{i+j}^T D_{i+j}^T A D_{i+j} \overset{d}{q}_{i+j} + \overset{d}{q}_{i+j}^T D_{i+j}^T A u_i - \overset{d}{q}_{i+j}^T D_{i+j}^T b + \sum_{j=1}^{n-1}\sum_{k=1}^{j} \overset{d}{q}_{i+j}^T D_{i+j}^T A D_{i+k-1} \overset{d}{q}_{i+k-1} + \mathcal{F}(u_i) \quad (30)$$

$$\mathcal{G}(u_{i+n}) = \sum_{j=0}^{n-1} \overset{d}{q}_{i+j}^T D_{i+j}^T A^T A D_{i+j} \overset{d}{q}_{i+j} + 2\overset{d}{q}_{i+j}^T D_{i+j}^T A^T A u_i - \overset{d}{q}_{i+j}^T D_{i+j}^T A^T b + \sum_{j=1}^{n-1}\sum_{k=1}^{j} \overset{d}{q}_{i+j}^T D_{i+j}^T A^T A D_{i+k-1} \overset{d}{q}_{i+k-1} + \mathcal{G}(u_i) \quad (31)$$

where $\mathcal{F}(\boldsymbol{u}_{i+n}), \mathcal{G}(\boldsymbol{u}_{i+n}): \mathbb{R}^N \to \mathbb{R}$ and $\overset{d}{\boldsymbol{q}}_i, \overset{d}{\boldsymbol{q}}_{i+1}, \ldots, \overset{d}{\boldsymbol{q}}_{i+n-1} \in \mathbb{R}^N$ each corresponding to $d^N$ eigenstates, collectively $d^{n \times N}$ eigenstates. Nested iterations allow $d^{n \times N}$ solutions of $\boldsymbol{u}_{i+n}$ corresponding to the eigenstates of $\left[\overset{d}{\boldsymbol{q}}_i, \overset{d}{\boldsymbol{q}}_{i+1}, \ldots, \overset{d}{\boldsymbol{q}}_{i+n-1}\right]^T$ to be collocated in $\boldsymbol{u} \in \mathbb{R}^N$ based on the choice of $\boldsymbol{D}_i, \boldsymbol{D}_{i+1}, \ldots, \boldsymbol{D}_{i+n-1}$ and $\boldsymbol{\Delta}_i, \boldsymbol{\Delta}_{i+1}, \ldots, \boldsymbol{\Delta}_{i+n-1}$. The Modified Ising Hamiltonian can then be used to anneal for either low-energy solutions or ground state solutions.

For $d = 2, n = 2$ $3^N$ and $4^N$ unique eigenstates can be collocated. This allows a modified Ising Hamiltonian problem of qutrits and ququarts to be formulated using qubits using the choices of $D_i$ and $\Delta_i$ provided in Table 1. Since current hardware is limited to qubits, for the remainder of this paper, we will work with the case $d = 2$ and denote $\overset{2}{\boldsymbol{q}}$ as $\boldsymbol{q}$ and use the fact that $\arg\min \tilde{E}\left(\overset{2}{\boldsymbol{q}}\right) = \arg\min E\left(\overset{2}{\boldsymbol{q}}\right)$ to formulate a modified Ising Hamiltonian problem and sample low energy eigenstates using a quantum annealer with the equivalent Ising Hamiltonian problem.

## 4. Hybrid direct search approach for quantum annealing

In Section 4.1, we provide a brief introduction to direct search and its properties. The cosine measures of some exponentially large positive spanning sets are presented. In Section 4.2, we present the iterative procedures used to perform a direct search using quantum annealers.

### 4.1. Direct Search

**Definition**: The positive span of a set of vectors $\{\boldsymbol{d}_1 \ldots \boldsymbol{d}_r\}$ in $\mathbb{R}^N$ is the convex cone

$$\{\boldsymbol{d} \in \mathbb{R}^N \mid \boldsymbol{d} = \alpha_1 \boldsymbol{d}_1 + \cdots \alpha_r \boldsymbol{d}_r \mid \alpha_i \geq 0 \mid 1 \leq i \leq r\} \tag{32}$$

**Definition**: A positive spanning set $D = \{\boldsymbol{d}_1 \ldots \boldsymbol{d}_r\}$ in $\mathbb{R}^N$ is a set of vectors that positively spans $\mathbb{R}^N$.

The set $D = \{\boldsymbol{d}_1 \ldots \boldsymbol{d}_r\}$ is said to be positively independent if none of the vectors is a positive linear combination of the others.

**Definition**: A positive basis in $\mathbb{R}^N$ is a positively independent set whose positive span is $\mathbb{R}^N$.

**Definition**: The cosine measure $cm(D)$ of a positive spanning set, $D$ is defined as

$$cm(D) = \min_{0 \neq \boldsymbol{v} \in \mathbb{R}^N} \max_{0 \neq \boldsymbol{d} \in D} \frac{\boldsymbol{v}^T \boldsymbol{d}}{\|\boldsymbol{v}\|_2 \|\boldsymbol{d}\|_2} \tag{33}$$

**Definition**: The simple maximal positive spanning basis $D_\oplus$ is defined as (Conn et al., 2009)

$$D_\oplus = \left\{ \boldsymbol{d}_\oplus^{(j)} \in \mathbb{R}^N \mid \boldsymbol{d}_{\oplus_i}^{(j)} = sgn(j)\delta_{i|j|} \mid -N \leq j \leq N, j \neq 0 \mid 1 \leq i \leq N \right\} \tag{34}$$

Classical direct search methods use $D_\oplus$ to sample $2N$ points $P_k = \{\boldsymbol{u}_k + \alpha_k \boldsymbol{d}_\oplus\}$ in the neighborhood of the current iterate $\boldsymbol{u}_k$ as shown in Figure 1 (a) to find a descent direction and search in a smaller neighborhood if a better solution is not found until some convergence criterion is met (Conn et al., 2009).

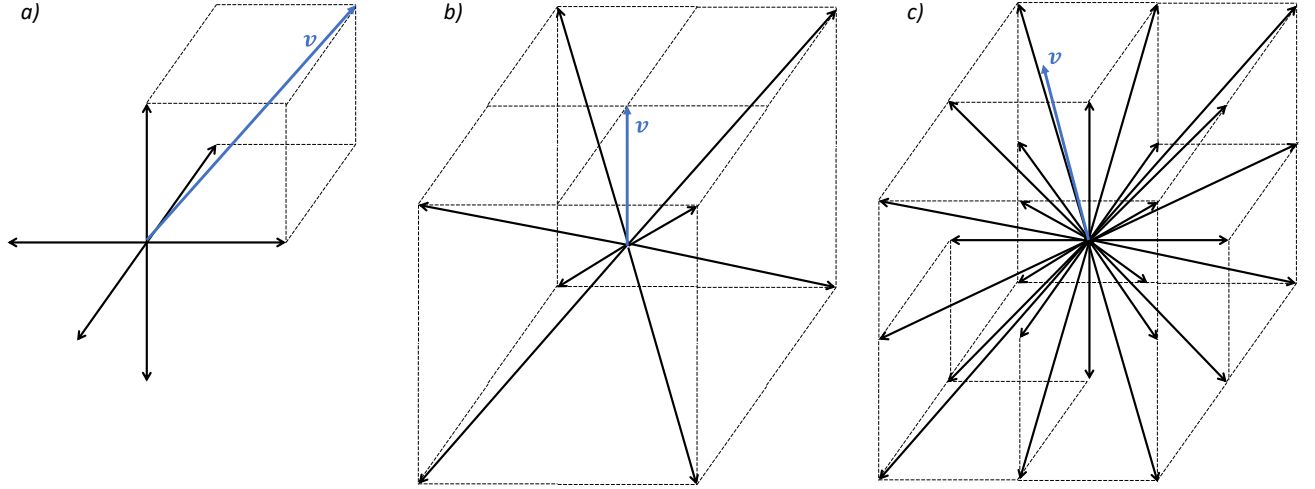

*Figure 1 The positive spanning sets (a) $D_\oplus$, (b) $D_2$, and (c) $D_3$, each shown with a corresponding minimizing vector $\boldsymbol{v}$ for $N = 3$*

The convergence of a direct search algorithm relies heavily on the cosine measure of the positive spanning set sampling points (Conn et al., 2009).

**Theorem 1**: (Conn et al., 2009) Let $D$ be a positive spanning set and some given $\alpha \in \mathbb{R}^+$. For a Lipschitz continuous $\nabla f$ with constant $v > 0$ in an open set containing the ball $B(x; \Delta)$ if $f(x) \leq f(x + \alpha d)\ \forall\ d \in D$ then

$$\|\nabla f(x)\| \leq \frac{v}{2} cm(D)^{-1} \max_{d \in D}\|d\|\alpha \tag{35}$$

Since $cm(D_\oplus) = \frac{1}{\sqrt{N}}$ (Nævdal, 2019), $cm(D_\oplus) \to 0$ with large $N$. In practical applications, convergence significantly slows down for $cm(D_\oplus) < 0.1$ (Conn et al., 2009) which is reached for $N = 100$. This makes classical direct search methods intractable for finite element problems exceeding a few hundred degrees of freedom. To overcome this problem, we examine the cosine measure of the positive spanning sets $D_2$ and $D_3$ formed by $N$ qubits and qutrits respectively.

**Definition**: The positive spanning set $D_2$ of N qubits is defined as

$$D_2 = \left\{\boldsymbol{d}_2^{(k)} \in \mathbb{R}^N\ \bigg|\ d_{2_l}^{(k)} = 2\left(\left[\boldsymbol{radix_2}(k)\right]_l - \frac{1}{2}\right)\bigg|\ 0 \leq k \leq 2^N - 1\ |\ 1 \leq l \leq N\right\} \tag{36}$$

where $\boldsymbol{radix_2}(k): \mathbb{R} \to \mathbb{R}^N$ is the base-2 representation of $k \in \mathbb{Z}^{nonneg}$ as a vector, of which the $l^{th}$ component is $[\boldsymbol{radix_2}(k)]_l$.

**Lemma 1**: $D_2$ is a positive spanning set in $\mathbb{R}^N$

**Proof**: Consider any $\boldsymbol{d}_\oplus^{(j)} \in D_\oplus$. It is always possible to find vectors $\boldsymbol{d}_{2_a}, \boldsymbol{d}_{2_b} \in D_2$ s.t.

$$\boldsymbol{d}_{2_a} = \begin{cases} \boldsymbol{d}_2^{(2^N - 1 - 2^{|j|})}, & j < 0 \\ \boldsymbol{d}_2^{(2^j)}, & j > 0 \end{cases} \tag{37}$$

$$d_{2_b} = \begin{cases} d_2^{(0)} & , j < 0 \\ d_2^{(2^N)} & , j > 0 \end{cases} \qquad (38)$$

to form the positive linear combination $d_\oplus^{(j)} = \frac{d_{2_a} + d_{2_b}}{2}$

**Theorem 2**: $cm(D_2) = \frac{1}{\sqrt{N}}$

**Proof**: Consider $0 \neq v \in \mathbb{R}^N$ s.t. $v_i = \|v\|_\infty$. For all such $i$, $v$ lies in a convex cone formed by $2^{N-1}$ vectors in $d_{2i}^{(k)} \in D_{2i} \subset D_2$ s.t.

$$D_{2i} = \left\{ d_{2i}^{(k)} \in \mathbb{R}^N \,\middle|\, d_{2i_l}^{(k)} = 2\left([radix_2(k)]_l - \frac{1}{2}\right) \middle| d_{2i_i}^{(k)} = sgn(v_i) \,\middle|\, 0 \leq k \leq 2^N - 1 \middle| 1 \leq l \leq N, l \neq i \right\} \qquad (39)$$

**Lemma 2**: $d_{2i}^{(k)} \in D_{2i}$ form the smallest $2^{N-1}$ angles for any $0 \neq v \in \mathbb{R}^N$

**Proof**: $\forall\, d_2^{(k)} \notin D_{2i}\; \exists\, d_{2i}^{(k+sgn(v_i)\times 2^i)} \in D_{2i}$ s.t. $d_2^{(k)} = d_{2i}^{(k+sgn(v_i)\times 2^i)} - sgn(v_i) \times e^{(i)}$ ∴
$\forall\, v^T d_2^{(k)} \;\exists\, v^T d_{2i}^{(k+sgn(v_i)\times 2^i)} \geq v^T d_2^{(k)} \;\because\; d_{2i_i}^{(k+sgn(v_i)\times 2^i)} = sgn(v_i)\; \forall\, 0 \leq k \leq 2^N - 1$ (40)

where $e^{(i)}$ is the $i^{th}$ standard basis vector

**Lemma 3**: The unique unit vector equidistant from all $d_{2i} \in D_{2i}$ and lying in the convex cone formed by $d_{2i} \in D_{2i}$ is $sgn(v_i) \times e^{(i)}$

**Proof**:

$$\frac{d_{2i}^T\, sgn(v_i)\times e^{(i)}}{\|d_{2i}\|_2 \|sgn(v_i)\times e^{(i)}\|_2} = \frac{1}{\sqrt{N}} \;\forall\, d_{2i} \in D_{2i}$$

Therefore, any other $v \neq d_\oplus^{sgn(v_i)}$ in the convex cone $D_{2i}$ has a smaller angle with some $d_{2i} \in D_{2i}$.

Using Lemmas 2 and 3, $cm(D_2) = \frac{1}{\sqrt{N}}$.

**Definition**: The positive spanning set $D_3$ of N qutrits is defined as

$$D_3 = \left\{ d_3^{(k)} \in \mathbb{R}^N \,\middle|\, d_{3_l}^{(k)} = [radix_3(k)]_l - 1 \middle| 0 \leq k \leq 3^N - 1 \,\middle|\, 1 \leq l \leq N \right\} \qquad (40)$$

where $radix_3(k): \mathbb{R} \to \mathbb{R}^N$ is the base-3 representation of $k \in \mathbb{Z}^{nonneg}$ as a vector, of which the $l^{th}$ component is $[radix_3(k)]_l$.

**Note**: $D_3$ positively spans $\mathbb{R}^N$ as $D_\oplus \subset D_3$

**Theorem 3**: $cm(D_3) = \dfrac{1}{\sqrt{\sum_{j=1}^N (\sqrt{j} - \sqrt{j-1})^2}}$ (41)

**Proof**: Consider $0 \neq v \in \mathbb{R}^N$ and a corresponding permutation matrix

$$P_v \in \mathbb{R}^{N \times N} \text{ s.t. } |[P_v v]_l| \geq |[P_v v]_{l+1}| \;\forall\, 1 \leq l \leq n-1 \qquad (42)$$

and the convex cone $D_{3v} \subset D_3$, where

$$D_{3v} = \left\{ \boldsymbol{d}_{3v}^{(m)} \in \mathbb{R}^N \mid \left[ \boldsymbol{P}_v^{-1} \boldsymbol{d}_{3v}^{(m)} \right]_l = sgn([\boldsymbol{P}^{-1}v]_l) \, H(m-l) \mid 1 < m < N \mid 1 < l < N \right\} \quad (43)$$

with $H(x) = \begin{cases} 1, & x \geq 0 \\ 0, & x < 0 \end{cases}$ \hfill (44)

$\boldsymbol{P}_v$ and $D_{3v}$ can be found for any $\boldsymbol{v}$. Also, $\boldsymbol{v}$ lies in the convex cone formed by $D_{3v}$.

We also define the set of unit vectors

$$\widehat{D}_{3v} = \left\{ \widehat{\boldsymbol{d}}_{3v}^{(m)} \in \widehat{D}_{3v} \mid \widehat{\boldsymbol{d}}_{3v}^{(m)} = \frac{\boldsymbol{d}_{3v}^{(m)}}{\sqrt{N-m}} \right\} \quad (45)$$

**Lemma 4**: $\boldsymbol{d}_{3v}^{(m)} \in D_{3v}$ form the smallest $N$ angles with $\boldsymbol{v}$ among all $\boldsymbol{d}_3^{(k)} \in D_3$.

**Proof**: Any $\boldsymbol{d}_3^{(k)} \notin D_{3v}$ satisfies the following property:

$$\boldsymbol{v}^T \boldsymbol{d}_3^{(k)} \leq \boldsymbol{v}^T \boldsymbol{d}_{3i}^{(m)} \quad \forall \, 0 < k < 3^N - 1, m = \left\| \boldsymbol{d}_3^{(k)} \right\|_1 \quad (46)$$

Therefore $\dfrac{\boldsymbol{v}^T \boldsymbol{d}_3^{(k)}}{\|\boldsymbol{v}\|_2 \|\boldsymbol{d}_3^{(k)}\|} \leq \dfrac{\boldsymbol{v}^T \boldsymbol{d}_{3v}^{(m)}}{\|\boldsymbol{v}\|_2 \|\boldsymbol{d}_{3v}^{(m)}\|} \quad \forall \, 0 < k < 3^N - 1, m = \left\| \boldsymbol{d}_3^{(k)} \right\|_1 \quad (47)$

**Lemma 5**: $\exists \, \boldsymbol{r}_v$ s.t. $[\boldsymbol{P}^{-1}\boldsymbol{r}_v]_i = r_1(\sqrt{i} - \sqrt{i-1})sign([\boldsymbol{P}^{-1}v]_i)$ lying in the convex cone formed by $D_{3v}$ which forms equal angles with all $\boldsymbol{d}_{3v} \in D_{3v}$ where $r_1 \in \mathbb{R}^+$, $1 \leq i \leq N$

**Proof**:

We prove the existence of $\boldsymbol{r}_v$ by construction.

Assume

$$\boldsymbol{P}^{-1}\boldsymbol{r}_v = (r_1, r_2, \ldots, r_n)^T = \boldsymbol{r}_{Pv} \quad (48)$$

Define

$$\widehat{\boldsymbol{r}}_{Pv} = \frac{\boldsymbol{r}_{Pv}}{\|\boldsymbol{r}_{Pv}\|_2} = \frac{1}{c}(r_1, r_2, \ldots, r_n)^T \quad (49)$$

where $c = \|\boldsymbol{r}_{Pv}\|_2$

In order to lie in the convex cone formed by $D_{3v}$, $r_1 \in \mathbb{R}^{sgn(v_i)}$

By definition

$$\frac{\widehat{\boldsymbol{r}}_{Pv}^T \widehat{\boldsymbol{d}}_{3v}^{(1)}}{\|\widehat{\boldsymbol{r}}_{Pv}\|_2 \|\widehat{\boldsymbol{d}}_{3v}^{(1)}\|_2} = \frac{\widehat{\boldsymbol{r}}_{Pv}^T \widehat{\boldsymbol{d}}_{3v}^{(2)}}{\|\widehat{\boldsymbol{r}}_{Pv}\|_2 \|\widehat{\boldsymbol{d}}_{3v}^{(2)}\|_2} \quad (50)$$

$$\frac{r_1}{c} = \frac{r_1 + r_2}{c\sqrt{2}} \quad (51)$$

$$r_2 = (\sqrt{2} - \sqrt{1})r_1 \quad (52)$$

Similarly

$$\frac{\hat{r}_{Pv}{}^T \hat{d}_{3v}^{(2)}}{\|\hat{r}_{Pv}\|_2 \|\hat{d}_{3v}^{(2)}\|_2} = \frac{\hat{r}_{Pv}{}^T \hat{d}_{3v}^{(3)}}{\|\hat{r}_{Pv}\|_2 \|\hat{d}_{3v}^{(3)}\|_2} \tag{53}$$

$$\frac{r_1+r_2}{\sqrt{2}c} = \frac{r_1+r_2+r_3}{\sqrt{3}c} \tag{54}$$

$$\frac{r_1+\sqrt{2}r_1-\sqrt{1}r_1}{\sqrt{2}c} = \frac{r_1+\sqrt{2}r_1-\sqrt{1}r_1+r_3}{\sqrt{3}c} \tag{55}$$

$$r_3 = (\sqrt{3}-\sqrt{2})r_1 \tag{56}$$

This leads to the general form

$$r_j = (\sqrt{j}-\sqrt{j-1})r_1 \ \forall \ 1 \leq j \leq N \tag{57}$$

and

$$c = \|\boldsymbol{r}_{Pv}\|_2 = \sqrt{\sum_{j=1}^{N}(\sqrt{j}-\sqrt{j-1})^2 r_1^2} = r_1\sqrt{\sum_{j=1}^{N}(\sqrt{j}-\sqrt{j-1})^2} \tag{58}$$

Thereby completing the unique determination of $\hat{r}_{Pv}$ independent of $r_1$

$$[\hat{r}_{Pv}]_j = \frac{\sqrt{j}-\sqrt{j-1}}{\sqrt{\sum_{j=1}^{N}(\sqrt{j}-\sqrt{j-1})^2}} \tag{59}$$

$$\boldsymbol{r}_{Pv} = r_1 \hat{r}_{Pv} \tag{60}$$

$$\boldsymbol{r}_v = \boldsymbol{P} \boldsymbol{r}_{pv} \tag{61}$$

∴ $\boldsymbol{r}_v$ unique to $\boldsymbol{P}$, can be constructed for any $\boldsymbol{v}$.

**Lemma 6**: $\dfrac{\boldsymbol{r}_v^T \boldsymbol{d}_{3v}^{(m)}}{\|\boldsymbol{r}_v^T\| \|\boldsymbol{d}_{3v}^{(m)}\|} = \dfrac{1}{\sqrt{\sum_{j=1}^{N}(\sqrt{j}-\sqrt{j-1})^2}} \quad \forall \ \boldsymbol{d}_{3v}^{(m)} \in D_{3v}, 1 \leq m \leq N$ (62)

**Proof**: By definition and construction

$$\frac{\boldsymbol{r}_v^T \boldsymbol{d}_{3v}^{(m)}}{\|\boldsymbol{r}_v\|_2 \|\boldsymbol{d}_{3v}^{(m)}\|_2} = \frac{\boldsymbol{r}_v^T \boldsymbol{d}_{3v}^{(1)}}{\|\boldsymbol{r}_v\|_2 \|\boldsymbol{d}_{3v}^{(1)}\|_2} = \frac{r_1}{r_1\sqrt{\sum_{j=1}^{N}(\sqrt{j}-\sqrt{j-1})^2}} = \frac{1}{\sqrt{\sum_{j=1}^{N}(\sqrt{j}-\sqrt{j-1})^2}} \ \forall \ 1 \leq m \leq N \tag{63}$$

Using Lemmas 5 and 6,

$$cm(D_3) = \frac{1}{\sqrt{\sum_{j=1}^{N}(\sqrt{j}-\sqrt{j-1})^2}} \tag{64}$$

**Theorem 4**: $cm(D_3) \geq \dfrac{1}{\sqrt{\ln(N)+\gamma+O\left(\frac{1}{N}\right)}} \geq \dfrac{1}{\sqrt{\ln(N)+1}} \ \forall \ N \geq 1$ (65)

where $\gamma$ is the Euler-Mascheroni constant (Euler, 1734).

**Proof**: Consider $\sum_{i=1}^{N}(\sqrt{j}-\sqrt{j-1})^2$

**Corollary 1**: $0 \leq (j-\sqrt{j^2-j})^2 \leq 1 \ \forall \ j \geq 1$ (66)

**Proof**:

We prove by assuming

$$0 \leq \left(j - \sqrt{j^2 - j}\right)^2 \leq 1 \;\forall\; j \geq 1 \tag{67}$$

$$j \geq \sqrt{j^2 - j} \geq j - 1$$

$$j^2 \geq j^2 - j \geq j^2 - 2j + 1$$

$$2j \geq j \geq 1 \tag{68}$$

which holds true $\forall\; j \geq 1 \therefore 0 \leq \left(j - \sqrt{j^2 - j}\right)^2 \leq 1 \;\forall\; j \geq 1$

Using Corollary 1

$$\sum_{j=1}^{N}\left(\sqrt{j} - \sqrt{j-1}\right)^2 = \sum_{j=1}^{N} \frac{\left(j - \sqrt{j^2 - j}\right)^2}{\left(\sqrt{j}\right)^2} \leq \sum_{j=1}^{N} \frac{1}{j} \;\forall\; N \geq 1 \tag{69}$$

$\sum_{j=1}^{N} \frac{1}{j}$ is the harmonic series which has the first order approximation (Euler, 1734, 1737)

$$\sum_{j=1}^{N} \frac{1}{j} = \ln(N) + \gamma + O\left(\frac{1}{N}\right) \tag{70}$$

$$\therefore \sum_{j=1}^{N}\left(\sqrt{j} - \sqrt{j-1}\right)^2 \leq \ln(N) + \gamma + O\left(\frac{1}{N}\right) \leq \ln(N) + 1 \tag{71}$$

$$\frac{1}{\sum_{j=1}^{N}\left(\sqrt{j} - \sqrt{j-1}\right)^2} \geq \frac{1}{\ln(N) + \gamma + O\left(\frac{1}{N}\right)} \geq \frac{1}{\ln(N) + 1} \tag{72}$$

$$\frac{1}{\sqrt{\sum_{j=1}^{N}\left(\sqrt{j} - \sqrt{j-1}\right)^2}} \geq \frac{1}{\sqrt{\ln(N) + \gamma + O\left(\frac{1}{N}\right)}} \geq \frac{1}{\sqrt{\ln(N) + 1}} \tag{73}$$

$$cm(D_3) \geq \frac{1}{\sqrt{\ln(N) + \gamma + O\left(\frac{1}{N}\right)}} \geq \frac{1}{\sqrt{\ln(N) + 1}} \tag{74}$$

**Definition**: The positive spanning set $D_4$ of N qutrits is defined as

$$D_4 = \left\{\boldsymbol{d}_4^{(k)} \in \mathbb{R}^N \;\middle|\; d_{4_l}^{(k)} = [\boldsymbol{radix_4}(k)]_l - 1 \;\middle|\; 0 \leq k \leq 4^N - 1 \;\middle|\; 1 \leq l \leq N \right\} \tag{75}$$

where $\boldsymbol{radix_4}(k): \mathbb{R} \to \mathbb{R}^N$ is the base-4 representation of $k \in \mathbb{Z}^{nonneg}$ as a vector, of which the $l^{th}$ component is $[\boldsymbol{radix_4}(k)]_l$.

**Note**: $D_4$ positively spans $\mathbb{R}^N$ as $D_2 \subset D_4$

**Conjecture 1**: $cm(D_4) \geq cm(D_3)$ (76)

While quantum annealing using $D_2$ does not yield any advantage over $D_\oplus$ in cosine measure, points in $D_3$ result in an exponential improvement, e.g., $cm(D_3) > 0.1$ for $N < 10^{42} < e^{99}$ using Theorem 4.

Another important aspect to consider in direct search is the length of the largest vector in the sampling points around the last iterate. Classical methods like pattern search typically use $D_\oplus$ and all vectors in $D_\oplus$ are of equal length and can only move in the coordinate directions, requiring $N$ iterations to move in the diagonal direction. Vectors in $D_2$ also have equal length but allow movement in the diagonal directions in a single iteration while requiring two iterations to move in any coordinate direction. Vectors in $D_3$ have a ratio of the maximum to the minimum length of $\sqrt{N}$ and allow movement in both the coordinate and diagonal directions in a single step. Hence, searches using $D_2$ and $D_3$ for $\boldsymbol{u} \in \mathbb{R}^N$ can cover significantly longer distances in $\mathbb{R}^N$ with a single iteration for large $N$.

## 4.2. Iterative Procedures

The performance of iterative direct search algorithms is dependent on the initial guess and the size of the initial search neighborhood. To mitigate these effects, we also introduce an expansion phase as demonstrated in Figure 2 (a) that increases the search neighborhood for the initial search until the iteration fails, after which the search is performed in the usual manner by decreasing the search neighborhood till the convergence criterion is met as shown in Figure 2 (b). The expansion phase avoids slow initial convergence arising from a poor selection of an initial guess or a small initial search neighborhood.

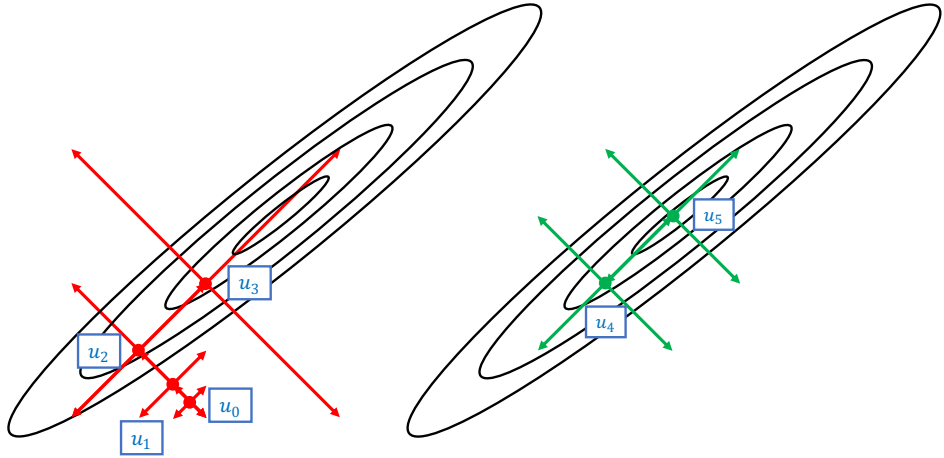

*Figure 2 Direct search demonstrating (a) the initial expansion phase (b) the subsequent contraction phase*

### 4.2.1. $2^N, 3^N, and\ 4^N$ Search

This iterative procedure can use the $D_2, D_3, D_4$ positive spanning sets as the sampling points. Since current quantum annealers are limited to qubits, $D_3, D_4$ can be generated using nested iterations with a choice of $\boldsymbol{\Delta}_i, \boldsymbol{\Delta}_{i+1}, \dots, \boldsymbol{\Delta}_{i+n} = \boldsymbol{0}$ and $\mathbf{D}_i, \mathbf{D}_{i+1}, \dots, \mathbf{D}_{i+n}$ as summarized in Table 1. $D_2, D_3, D_4$ are shown in $\mathbb{R}^2$ in Figure 3.

*Table 1 Parameters for nested iterations to create $D_2, D_3, D_4$*

| | $\mathbf{D}_i$ | $\boldsymbol{\Delta}_i$ |
|---|---|---|
| $D_2$ | | $\boldsymbol{\Delta}_i = \boldsymbol{0}$ |
| $D_3$ | $\mathbf{D}_{i+1} = \mathbf{D}_i$ | $\boldsymbol{\Delta}_i = \boldsymbol{\Delta}_{i+1} = \boldsymbol{0}$ |
| $D_4$ | $\mathbf{D}_{i+1} = \mathbf{D}_i/2$ | $\boldsymbol{\Delta}_i = \boldsymbol{\Delta}_{i+1} = \boldsymbol{0}$ |

$D_3$ using nested iterations corresponds to $4^N$ eigenstates of $\{\overset{2}{\boldsymbol{q}}_i, \overset{2}{\boldsymbol{q}}_{i+1}\}$ but due to overlap of the collocated points, redundant eigenstates lead to $3^N$ unique $\boldsymbol{u}_{i+2}$ as shown in Figure 3 (b).

$u_{i+1}^{(j)} = u_i + \alpha_i d \ni d \in D_2, D_3, D_4$ are sought by sampling from annealing such that $\mathcal{F}(u_{i+1}) < \mathcal{F}(u_i)$ in a polling step and $u_{best} = \arg\min_j \mathcal{F}\left(u_{i+1}^{(j)}\right)$ is chosen. If such $u_{best}$ is found polling is successful and $\alpha_{i+1}$ is chosen to be $\alpha_i$ or $2\alpha_i$ and the next iteration polls around $u_{i+1} = u_{best}$. If $\mathcal{F}(u_{best}) < \mathcal{F}(u_i)$ is not found polling is unsuccessful and $\alpha_{i+1} = \alpha_i/2$ and the next iteration polls around $u_i$.

The overlap in $D_3$ skews the probability distribution of the resulting $u_{i+2}$ away from the diagonal directions towards the coordinated directions due to the redundancies as shown in Figure 3 (b). The larger number of eigenstates for nested iterations may require more samples from the annealing to obtain satisfactory performance.

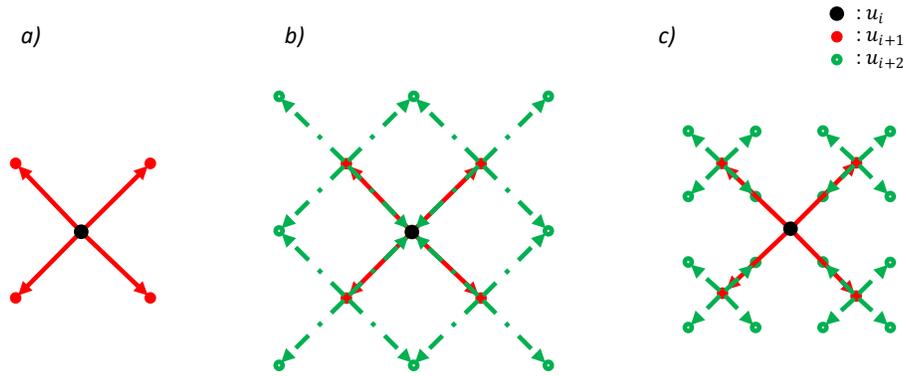

*Figure 3 a) $2^N$ search b) $3^N$ search using nested iterations with overlapping solutions b) $4^N$ search using nested iterations*

Algorithm 2. Pseudo-code of $2^N, 3^N, and\ 4^N$ Search

1. **Input** $A, b, \Psi_{min}, G, j_{max}, \mathcal{L}(u) = \mathcal{F}(u)\ or\ \mathcal{G}(u)$ **Initialization:** $u_0, \alpha_0, \Psi_{min} > 0$
2. **while** $\Psi(u_i) > \Psi_{min}, i = 0,1,2, ...$
3.     **Poll Step**
4.        Calculate $J_i, h_i$
5.        **for** $j = 1,2, ..., j_{max}$
6.           $q_i^{(j)} = anneal(J_i, h_i)$
7.        **end for**
8.        $u_{best} = \arg\min_j \mathcal{L}\left(u_i + D_i q_i^{(j)}\right)$
9.        **if** $\mathcal{L}(u_{best}) < \mathcal{L}(u_i)$
10.           $u_{i+1} = u_{best}$    Poll Successful
11.        **else**
12.           $u_{i+1} = u_i$    Poll Unsuccesful
13.        **end if**
14.     **Parameter Update**
15.        **if** Poll Successful
16.           $\alpha_{i+1} = \alpha_i$ (or $2\alpha_i$)
17.        **else**

| 18. | $\alpha_{i+1} = \alpha_i/2$ |
| 19. | **end if** |
| 20. | **end while** |

### 4.2.2. Hyperoctant Search

Hyperoctant search uses a two-step polling procedure using $D_2$. In the first step of polling $u_a^{(j)} = u_i + \alpha_i d \ni d \in D_2$ are sought by sampling from annealing and $u_a^{(j)} = \arg\min_j \mathcal{F}\left(u_a^{(j)}\right)$ is chosen. In the second step of polling $u_b^{(k)} = u_i + \Delta_i + \frac{\alpha_i}{2} d \ni d \in D_2$ and $\Delta_i = \frac{u_i + u_a}{2}$ are sought by sampling from the annealer. If any $u_a^{(j)}, u_b^{(k)}$ are found s.t. $\mathcal{F}\left(u_a^{(j)}\right) \leq \mathcal{F}(u_i)$ or $\mathcal{F}\left(u_b^{(k)}\right) \leq \mathcal{F}(u_i)$ then $u_{best} = \arg\min\left[\mathcal{F}\left(u_a^{(j)}\right), \mathcal{F}\left(u_b^{(k)}\right)\right]$ is chosen. If such $u_{best}$ is found, polling is successful and $\alpha_{i+1}$ is chosen to be $\alpha_i$ or $2\alpha_i$ and the next iteration polls around $u_{i+1} = u_{best}$. If $\mathcal{F}(u_{best}) < \mathcal{F}(u_i)$ is not found, polling is unsuccessful and $\alpha_{i+1} = \alpha_i/2$ and the next iteration polls around $u_i$.

**Note:** The cosine measure of hyperoctant search is bounded above and below by $cm(D_3)$ and $cm(D_2)$. If the minimum among $3^N$ grid points in the neighborhood of $u_i$ is $u_{min}$, in the best case $u_i + \frac{u_a}{2} + u_b = u_{min}$ and in the worst case $u_b = \frac{u_a}{2}$ when $u_i + \frac{u_a}{2} + u_b \neq u_{min}$.

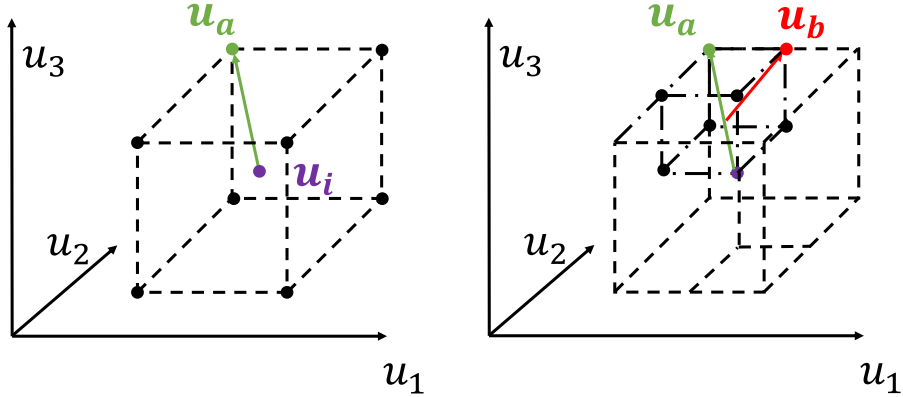

*Figure 4 The iterative procedure shown for $n = 3$ (a) The first step of an iteration: finding the best hyperoctant using $u_a$ (b) The second step of an iteration: probing the hyperoctant corresponding to $u_a$ for a better solution $u_b$.*

Algorithm 3. Pseudo-code of Hyperoctant Search

1. **Input $A, b, \Psi_{min}, G, j_{max}, \mathcal{L}(u) = \mathcal{F}(u)$ or $\mathcal{G}(u)$ Initialization: $u_0, \alpha_0, \Psi_{min} > 0$**
2. **while $\Psi(u_i) > \Psi_{min}, i = 0,1,2,...$**
3.    **Poll Step**
4.       Calculate $J_{i_a}, h_{i_a}$
5.       **for** $j = 1,2,...,j_{max}$
6.          $q_{i_a}^{(j)} = anneal(J_{i_a}, h_{i_a})$
7.       **end for**

8.     $u_a = \arg\min_j \mathcal{L}\left(u_i + D_{i_a}q_i^{(j)}\right)$
9.     Calculate $J_{i_b}, h_{i_b}$
10.    **for** $k = 1,2,\ldots,k_{max}$
11.        $q_{i_b}^{(k)} = anneal(J_{i_b}, h_{i_b})$
12.    **end for**
13.    $u_b = \arg\min_j \mathcal{L}\left(u_a + \Delta_{i_b} + D_{i_b}q_i^{(j)}\right)$
14.    $u_{best} = \arg\min[\mathcal{L}(u_a), \mathcal{L}(u_b)]$
15.    **if** $\mathcal{L}(u_{best}) < \mathcal{L}(u_i)$
16.        $u_{i+1} = u_{best}$    Poll Successful
17.    **else**
18.        $u_{i+1} = u_i$    Poll Unsuccesful
19.    **end if**
20.    **Parameter Update**
21.    **if** Poll Successful
22.        $\alpha_{i+1} = \alpha_i$ (or $2\alpha_i$)
23.    **else**
24.        $\alpha_{i+1} = \alpha_i/2$
25.    **end if**
26. **end while**

### 4.2.3. Benchmarking annealing using the time-to-target metric

Ground state solutions require exponentially increasing times with problem size for both classical and quantum techniques. Noise and discretization errors may prohibit them from being computed on quantum annealers. To circumvent these limitations, algorithms that do not rely on ground states may instead work with low eigenenergy eigenstates. Such algorithms can be benchmarked using the time-to-target ($TTT$) metric, which compares quantum annealing with classical techniques (King et al., 2015).

The $TTT$ metric uses a quantum annealer to sample low energy eigenstates of an Ising Hamiltonian from a quantum annealer. The "target" energy is identified using a choice of the $q^{th}$ percentile of the eigenenergy distribution from the samples. Similarly, a distribution of energies can be obtained using simulated annealing. The distributions are used to obtain the probabilities $\hat{p}_{t_{qa}}, \hat{p}_{t_{sa}}$ of obtaining a sample with energy lower than or equal to the target energy set by quantum annealing. Using these probabilities, the expected number of samples to obtain a sample with energy lower than or equal to the target energy can be obtained as

$$STT_{qa} = \frac{1}{\hat{p}_{t_{qa}}} \tag{77}$$

$$STT_{sa} = \frac{1}{\hat{p}_{t_{sa}}} \tag{78}$$

The annealing time $t_a$ for D-Wave devices can be chosen between $0.5 - 2000 \ \mu s$. The annealing time for simulated annealing is obtained by interpolating the best sweep time $t_s$ reported by (Borle & Lomonaco,

2018) for their optimized code and annealing using several choices of the number of sweeps $k_{max}$ choosing the optimum $TTT_{sa}$.

For D-Wave devices $t_a > 20\mu s$ is empirically known to provide little benefit for current hardware (King et al., 2015). Therefore $t_a = 20\mu s$ is selected. $STT_{qa}$ is obtained using 1000 samples, and the 10$^{th}$ percentile is chosen since 10 anneals are used for every Ising Hamiltonian for the results presented in Section 5. $STT_{sa}$ are obtained using 1000 samples for 10, 20, 40, 100, 200, 400, 1000, 2000, 4000, and 10000 sweeps. Using Equations (79) and (80) $TTT_{qa_{anneal}}$ and $TTT_{sa_{anneal}}$ are obtained and the lowest $TTT_{anneal_{sa}}$ among different sweep sizes is selected.

$$TTT_{qa_{anneal}} = STT_{qa}(t_a) \tag{79}$$

$$TTT_{sa_{anneal}} = STT_{sa} \times (k_{max} \times t_s) \tag{80}$$

The total wall clock time includes initialization and readout times for both simulated and quantum annealing. A detailed explanation of these may be found in (King et al., 2015).

## 5. Numerical examples

We present simulation results for problems in one- and two-dimensions to demonstrate the convergence behavior of the proposed iterative algorithm.

### 5.1. Poisson's equation in 1D

We consider the following Poisson's problem in 1D with homogeneous Dirichlet boundary conditions:

$Find\ u: \bar{\Omega} \to \mathbb{R}\ such\ that$

$$\nabla^2 u - f = 0\ in\ \Omega \in (0, L) \tag{81}$$

with $u(0) = 0; u(L) = 0$ and $f = -128\frac{x}{L} + 64$

The problem is discretized using linear finite elements. We compare different iterative procedures using simulated annealing since $3^N$ and $4^N$ search using nested iterations cannot be performed for large N on current hardware and the results are presented in Figure 5. Convergence with $4^N$ search and hyperoctant search is comparable. The poor convergence of $3^N$ Search is clearly seen due to the overlap in $D_3$ arising from nested iterations. Since hyperoctant search clearly converges faster while using fewer qubits and can be performed on quantum annealers for large N it is used for all subsequent results.

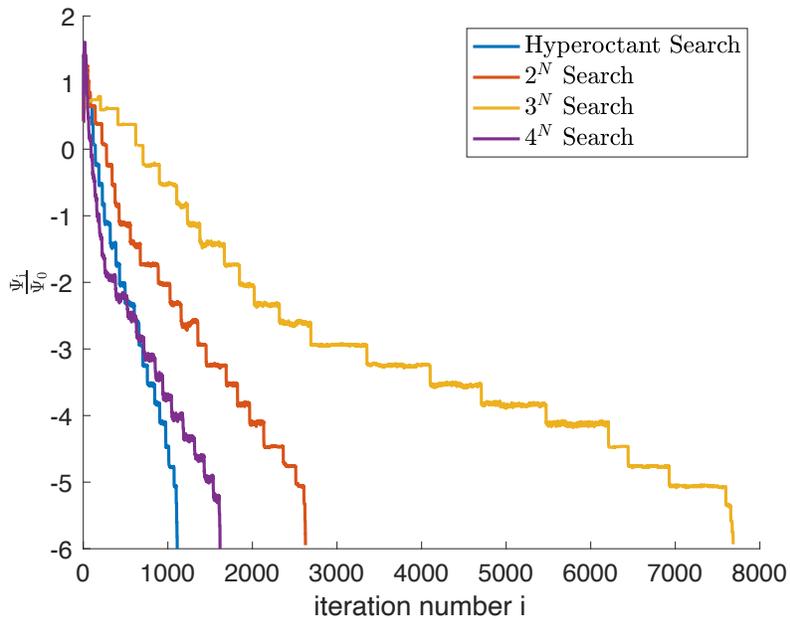

*Figure 5 Convergence of the residual of Hyperoctant Search, $2^N$ Search, $3^N$ Search, and $4^N$ Search using simulated annealing of the 1D Poisson's problem for N = 200*

We compare results using quantum annealing with classical simulated annealing. The D-Wave 2000Q is used for quantum annealing, and the D-Wave Ocean library is used for simulated annealing, each with ten 20 $\mu s$ anneals per iteration with 1 spin reversal transform used for quantum annealing and the optimum number of sweeps determined by the D-Wave Ocean library used for simulated annealing. The initial guess for the algorithm is set as zeros except for the nodes associated with Dirichlet boundary conditions. Figure 6 shows that both schemes exhibit similar convergence behavior when an h-refinement is performed.

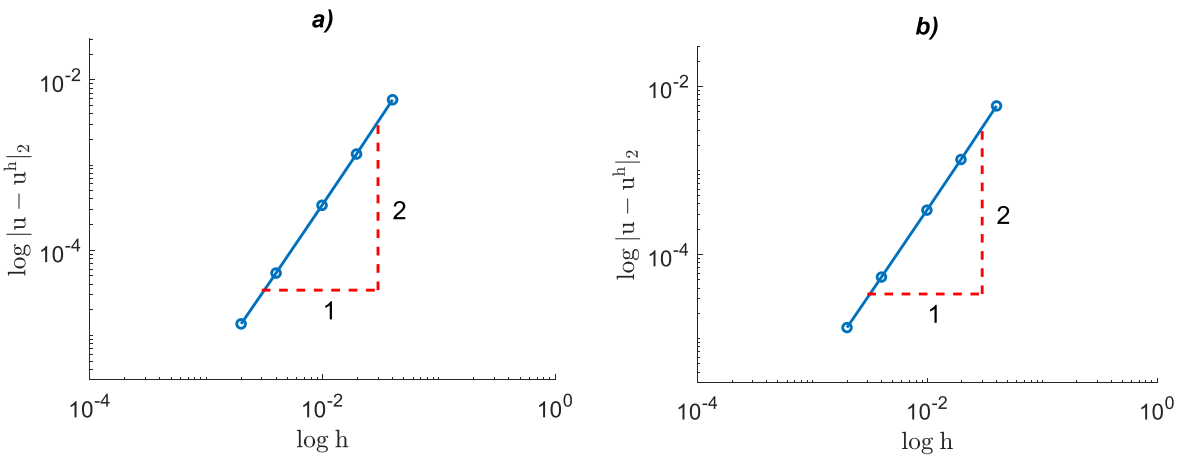

*Figure 6 Convergence in the $L_2$ norm of the solution of the 1D Poisson's problem obtained using a) quantum annealing b) simulated annealing when an h-refinement is performed.*

Figures 7 and 8 show the convergence of the functional $\mathcal{F}(\boldsymbol{u_i})$ being minimized iteratively and the residual in the $L_2$ norm of the linear system at each iteration. $\mathcal{F}(\boldsymbol{u_i})$ decreases monotonically, but flat regions and spikes can occasionally be seen for the residual. The level surfaces of $F(\boldsymbol{u_i})$ are ellipsoids, but the level surfaces of the residual are concentric spheroids and therefore a decrease in $\mathcal{F}(\boldsymbol{u_i})$ may not coincide with a decrease in the residual. The number of iterations is plotted instead of wall-clock time to compare the efficacy of quantum annealing over simulated annealing for the hyperoctant search iterative procedure, and quantum annealing shows a slight advantage over simulated annealing for the number of iterations.

Figure 9 shows the time to target metrics for simulated and quantum annealing. The energy distributions for quantum and simulated annealing are obtained using the D-Wave 2000Q with default settings and the D-Wave Ocean Library with varying numbers of simulated annealing sweeps, respectively, using 1000 anneals each. 10[th] percentile TTT metrics are calculated since 10 anneals were used per iteration. The final metrics are calculated by averaging the TTT metrics for 40 random Ising Hamiltonians encountered for 20 randomly selected iteration numbers for both steps of hyperoctant search.

Figure 9 clearly shows the increasing time overhead for simulated annealing while the time overhead for quantum annealing remains roughly constant for larger problems. The quantum tunneling property allows quantum annealing to tunnel out of local minima to explore regions with a large Hamming distance from the current iterate (Crosson & Harrow, 2016). Since quantum states are complex till they are measured, the exploration in this solution space is less restricted than simulated annealing. The properties of quantum entanglement eliminate the communication overhead required for simulated annealing in the form of repeated evaluations of the cost function. The trends match those reported by (King et al., 2015).

Although both quantum and simulated annealing converge in roughly the same order of iterations, a clear advantage in total annealing wall-clock time is seen for quantum annealing. Figure 10 shows a comparison of the estimated total wall-clock annealing time for simulated and quantum annealing to reach a normalized residual of $10^{-5}$.

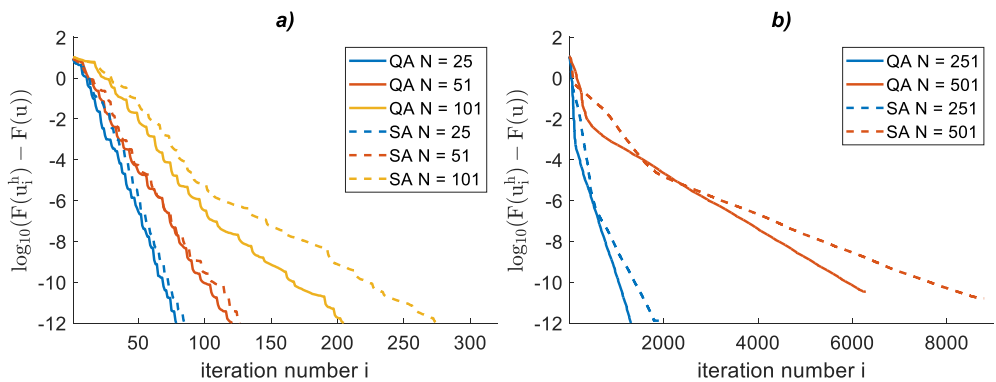

*Figure 7 Convergence of the functional $\mathcal{F}(\boldsymbol{u_i})$ plotted against the iteration number i for a) N = 25, 51, 101 and b) N = 251, 501. QA and SA stand for quantum annealing and simulated annealing, respectively.*

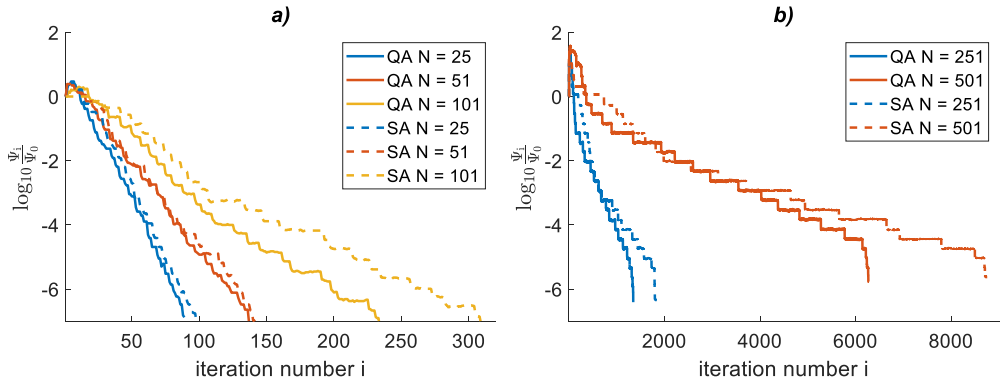

*Figure 8 Convergence of the normalized residual of the system of equations in the $L_2$ norm plotted against the iteration number i for a) N = 25, 51, 101 and b) N = 251, 501. QA and SA stand for quantum annealing and simulated annealing, respectively.*

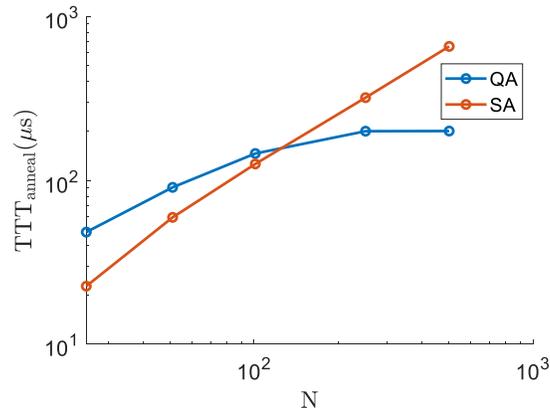

*Figure 9 Time to Target Metrics for the 1D Poisson's problem with both quantum and simulated annealing. N denotes the total number of nodes*

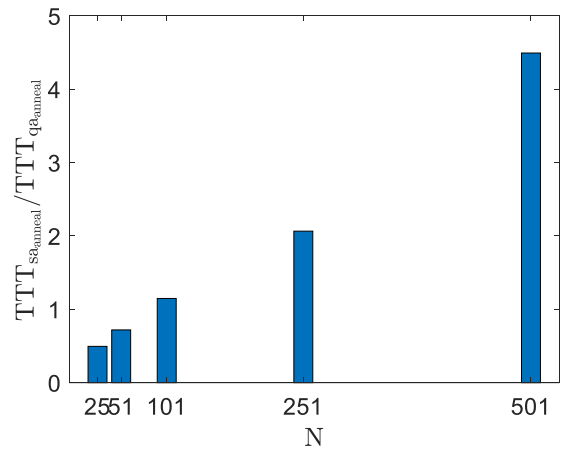

*Figure 10 Ratio of total annealing times for simulated and quantum annealing to solve to a precision of $\left|\frac{\Psi_i}{\Psi_0}\right| \leq 10^{-5}$*

### 5.2. 1D Wave Equation

We consider the following wave equation in 1D

$Find\ u: \overline{\Omega} \to \mathbb{R}\ such\ that$
$\ddot{u} - c^2 \nabla^2 u = 0\ in\ \Omega \in (0, L)$  (82)

where $c$ is a constant for the following initial and boundary conditions

| Case 1 | Case 2 | Case 3 | Case 4 | Case 5 |
|---|---|---|---|---|
| $u\left(0, \frac{x}{L}\right) = \sin\left(\frac{\pi x}{L}\right)$ $\dot{u}\left(0, \frac{x}{L}\right) = 0$ $u(t, 0) = 0$ $u(t, L) = 0$ | $u\left(0, \frac{x}{L}\right) = \sin\left(\frac{2\pi x}{L}\right)$ $\dot{u}\left(0, \frac{x}{L}\right) = 0$ $u(t, 0) = 0$ $u(t, L) = 0$ | $u\left(0, \frac{x}{L}\right) = \sin\left(\frac{25\pi x}{L}\right)$ $\dot{u}\left(0, \frac{x}{L}\right) = 0$ $u(t, 0) = 0$ $u(t, L) = 0$ | $u\left(0, \frac{x}{L}\right) = \sin\left(\frac{\pi x}{2L}\right)$ $\dot{u}\left(0, \frac{x}{L}\right) = 0$ $u(t, 0) = 0$ $\dot{u}(t, L) = 0$ | $u\left(0, \frac{x}{L}\right) = \sin\left(\frac{50\pi x}{2L}\right)$ $\dot{u}\left(0, \frac{x}{L}\right) = 0$ $u(t, 0) = 0$ $\dot{u}(t, L) = 0$ |

The problem is approximated using the method of weighted residuals and discretized using the finite element method with piecewise linear interpolating functions and h-refinement. We compare results using quantum annealing with classical simulated annealing. The D-Wave 2000Q is used for quantum annealing, and the D-Wave Ocean library is used for simulated annealing, each with 10 20 $\mu s$ anneals anneals per iteration with 1 spin reversal transform used for quantum annealing and the optimum number of sweeps determined by the D-Wave Ocean library used for simulated annealing. The initial guess for the algorithm is set as zeros. The Newmark method is used for implicit time-stepping with lumped mass matrices, and each time step is solved to a precision of $\|\Psi(u_i)\|_2 < 10^{-4}$ to economize annealer time. Solutions obtained using quantum annealing for four timesteps are shown in Figure 11.

Figure 12 and Figure 13 show the convergence of the functional $\mathcal{F}(u_i)$ being minimized iteratively and the residual in the $L_2$ norm of the linear system at each iteration. $\mathcal{F}(u_i)$ decreases monotonically, and again flat regions and spikes can occasionally be seen for the residual. The number of iterations is plotted instead of wall clock time to compare the efficacy of quantum annealing over simulated annealing for the hyperoctant search iterative procedure, and both techniques lead to convergence in roughly the same number of iterations. The number of iterations does not grow significantly with the problem size compared to Poisson's problem in 1D. Superconvergence is observed for Case 5. This is can also be explained by a near-exact solution coinciding with one of the collocated solutions in the search space and an anneal successfully returning it. The number of iterations is observed to increase for the higher harmonics. This can be explained by the fact that the solutions are further apart in the $L_2$ norm since acceleration values are higher for higher harmonics, and the Newmark method solves for accelerations, and corrections are made to obtain displacement and velocity.

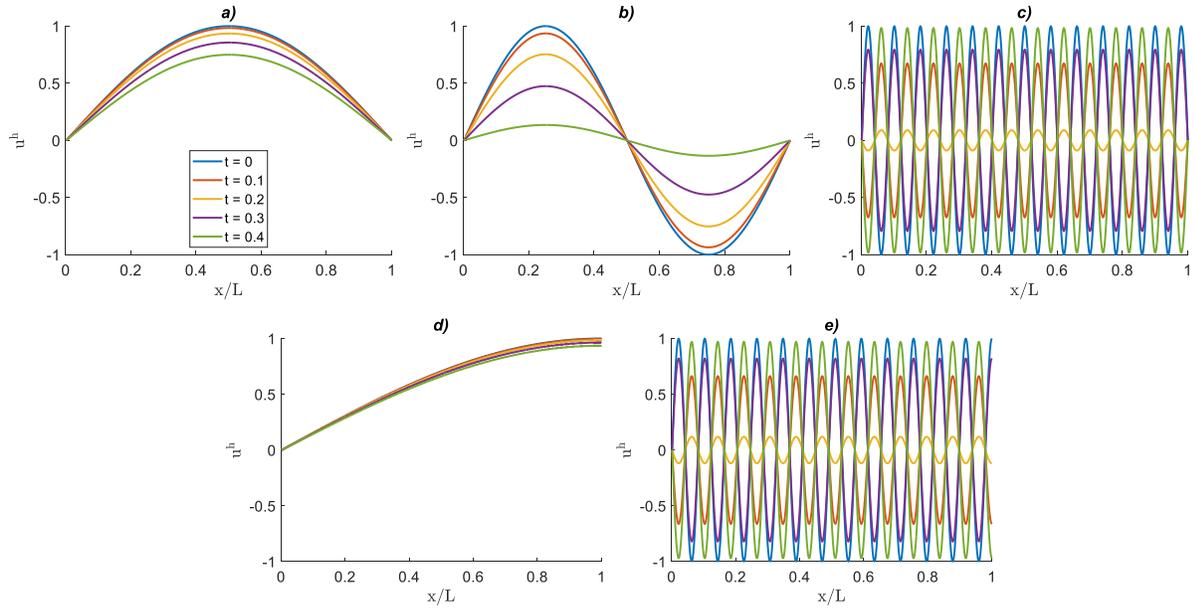

*Figure 11 Plot of the initial condition along with the four subsequent timesteps for (a) Case 1 (b) Case 2, (c) Case 3, (d) Case 4, (e) Case 5*

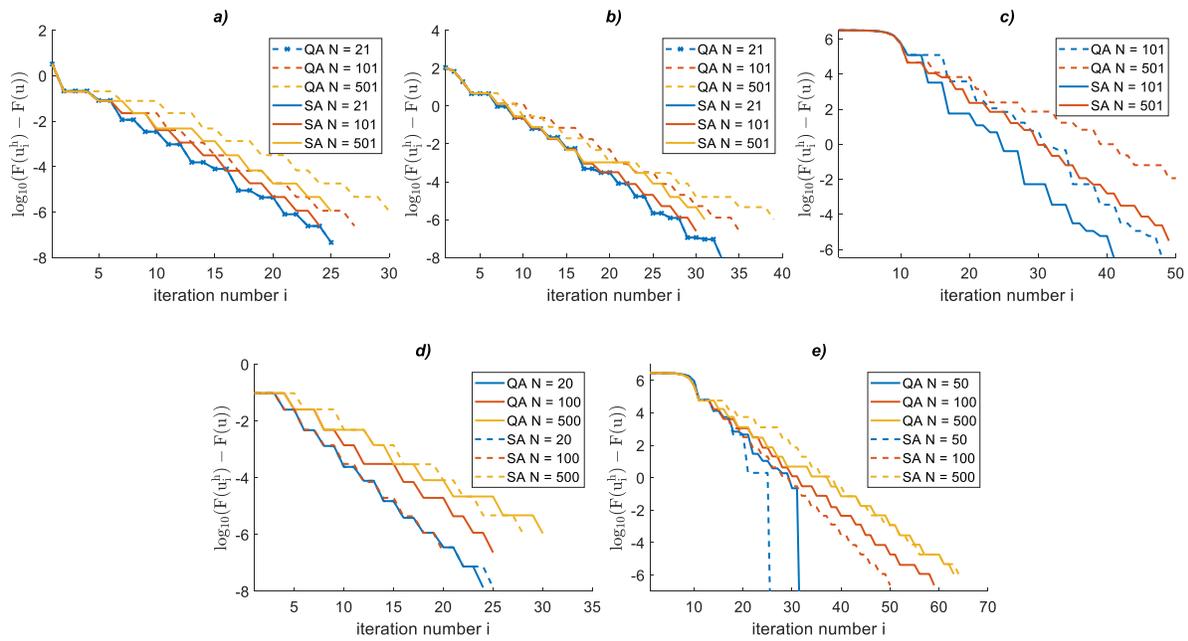

*Figure 12 Convergence of $\mathcal{F}(u_i)$ for the wave equation with both ends fixed shown for one timestep for (a) Case 1 (b) Case 2, (c) Case 3, (d) Case 4, (e) Case 5*

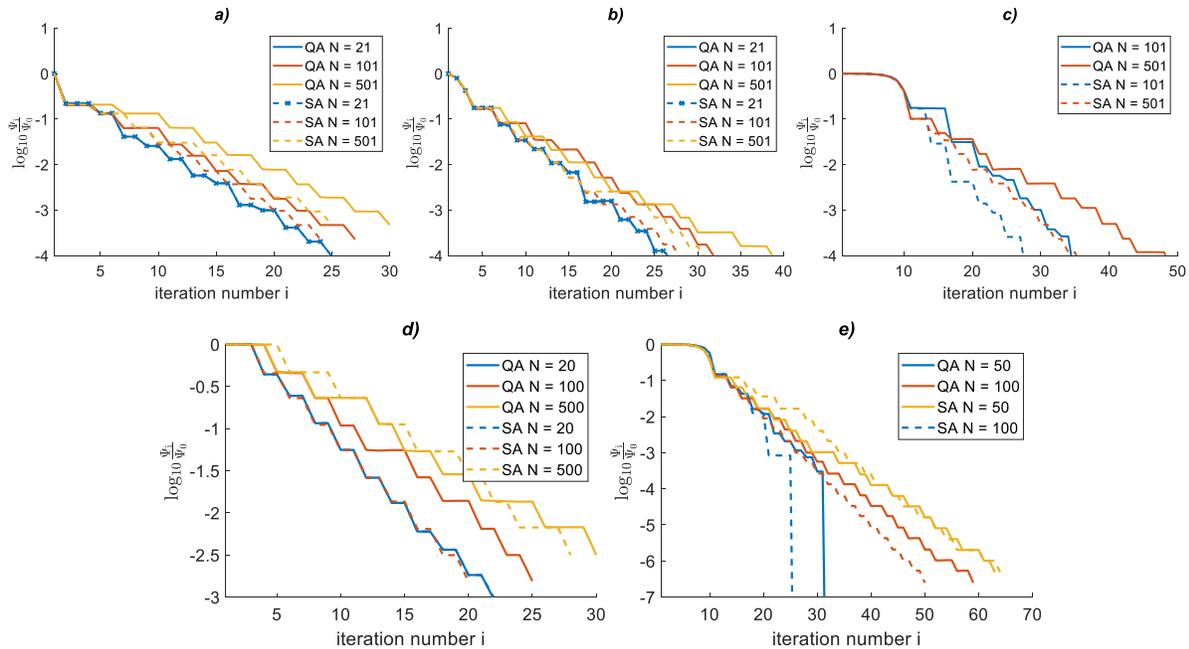

*Figure 13 Convergence of the residual for the wave equation with both ends fixed shown for one timestep for (a) Case 1 (b) Case 2, (c) Case 3, (d) Case 4, (e) Case 5*

Figure 14 shows the time to target metrics for both simulated and quantum annealing. The energy distributions for quantum annealing and simulated annealing are obtained using the D-Wave 2000Q with default settings and the D-Wave Ocean Library with varying numbers of simulated annealing sweeps, respectively, using 1000 anneals each. 10[th] percentile TTT metrics are calculated since 10 anneals were used per iteration. The final metrics are calculated by averaging the TTT metrics for 40 random Ising Hamiltonians encountered for 20 randomly selected iteration numbers for both steps of hyperoctant search. Although both quantum and simulated annealing converge in roughly the same order of iterations, a clear advantage in total annealing wall-clock time is seen for quantum annealing.

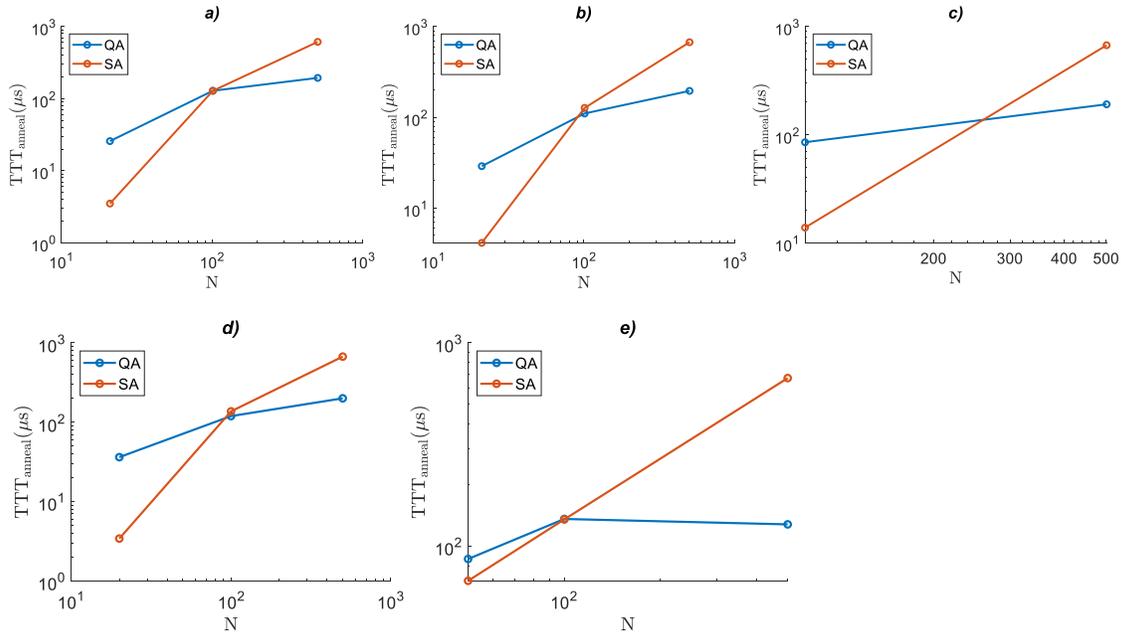

*Figure 14 Time to Target metrics for the wave equation, both ends fixed for (a) Case 1 (b) Case 2, (c) Case 3, (d) Case 4, (e) Case 5*

Figure 15 shows a comparison of the estimated total wall-clock annealing time for simulated and quantum annealing to reach a residual of $10^{-4}$.

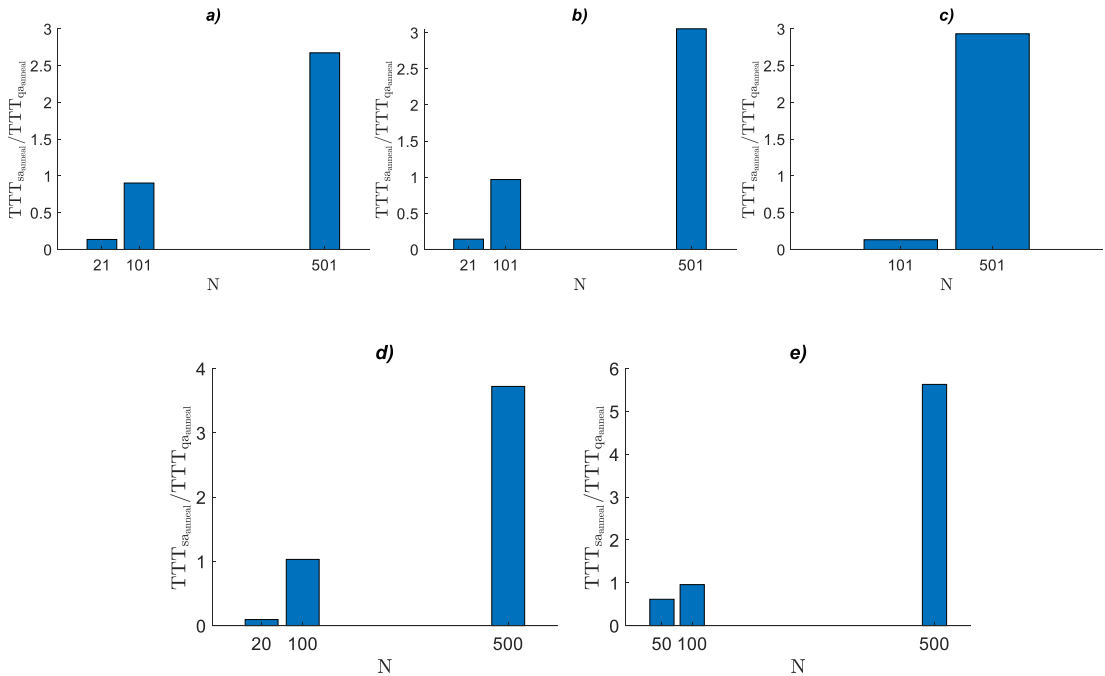

*Figure 15 Ratio of total annealing times for simulated and quantum annealing to solve to a precision of $\|\Psi(\mathbf{u}_i)\|_2 < 10^{-4}$ for (a) Case 1 (b) Case 2, (c) Case 3, (d) Case 4, (e) Case 5*

## 5.3. Poisson's problem in 2D

We consider the following Poisson's equation on a square domain:

Find $u: \bar{\Omega}^2 \to \mathbb{R}$ such that

$$\nabla^2 u - f = 0 \text{ in } \Omega \in (0, L) \times (0, L) \tag{83}$$

subject to

$$u = g \text{ on } \Gamma_g$$

$$-q_i n_i = h \text{ on } \Gamma_h$$

where $q_i = \kappa_{ij} u_{,j}$ and $n_i$ is the unit outward normal to $\Gamma = \Gamma_g \cup \Gamma_h$ and $\Gamma_g \cap \Gamma_h = \emptyset$.

The problem is approximated using the method of weighted residuals and discretized using the finite element method and four-noded bilinear elements. We compare results using quantum annealing with classical simulated annealing. The D-Wave Advantage and the D-Wave 2000Q is used for quantum annealing for Case 1 and Case 2, respectively, and the D-Wave Ocean library is used for simulated annealing, each with ten 20 $\mu s$ anneals per iteration with 1 spin reversal transform used for quantum annealing and the optimum number of sweeps determined by the D-Wave Ocean library used for simulated annealing. The initial guess for the algorithm is set as zeros except for the nodes associated with Dirichlet boundary conditions.

We consider two cases:

Case 1: $g = \frac{4}{L^2} x^2 - \frac{4}{L} x$ on $y = L$ and $y = 0$, $h = 0$ for $x = 0$ and $x = L$, and $f = 0.02$

Case 2: $g = 1$ on $y = L$ and $g = 0$ on $x = 0, x = L, y = 0$. $\Gamma_h = \emptyset$, and $f = 0$.

Figure 16 shows the solutions obtained using quantum annealing for a mesh of $15 \times 15$ nodes for both cases.

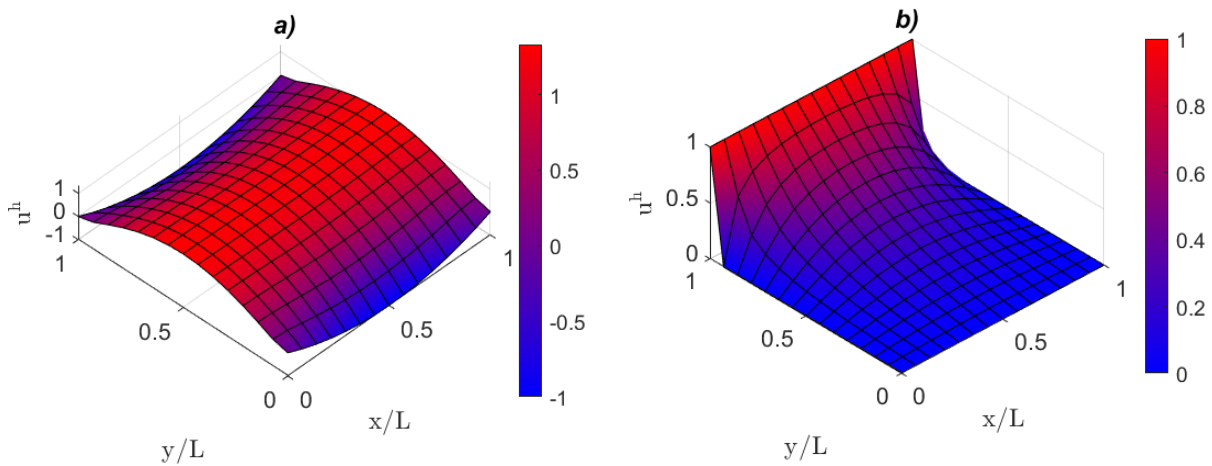

*Figure 16 Solution on a mesh of 15×15 nodes for a) Case 1 b) Case 2*

Figures 17 and 18 show the convergence of the functional $\mathcal{F}(\boldsymbol{u_i})$ being minimized iteratively and the residual in the $L_2$ norm of the linear system at each iteration. $\mathcal{F}(\boldsymbol{u_i})$ decreases monotonically, and again flat regions and spikes can occasionally be seen for the residual. The number of iterations is plotted instead of wall clock time to compare the efficacy of quantum annealing over simulated annealing for the hyperoctant search iterative procedure, and both techniques lead to convergence in roughly the same number of iterations, with quantum annealing demonstrating a slight advantage for larger problem sizes. The number of iterations does not grow significantly with the problem size compared to Poisson's problem in 1D, and the number of iterations is lower compared to Poisson's problem in 1D when problems with similar N are considered.

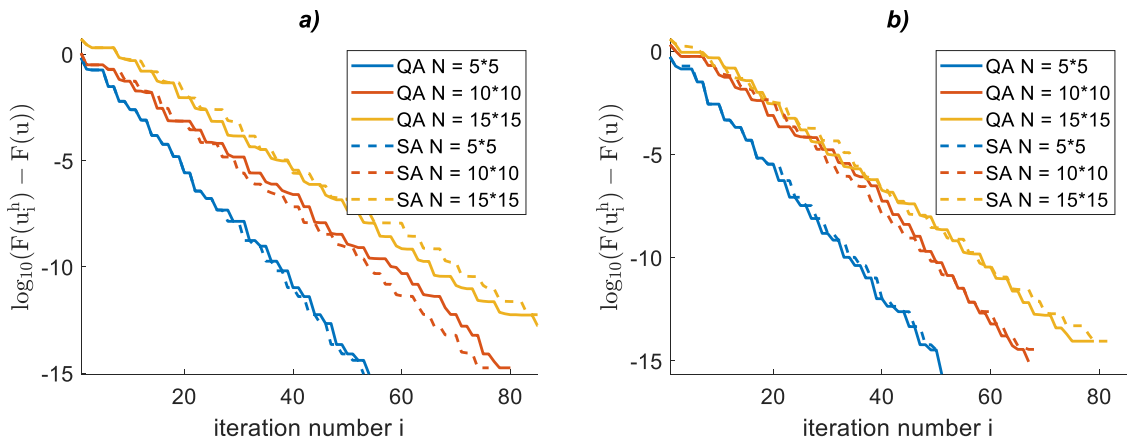

Figure 17 Convergence of the functional $\mathcal{F}(\boldsymbol{u_i})$ plotted against iteration for mesh of $5 \times 5, 10 \times 10,$ and $15 \times 15$ nodes for a) Case 1 and b) Case 2

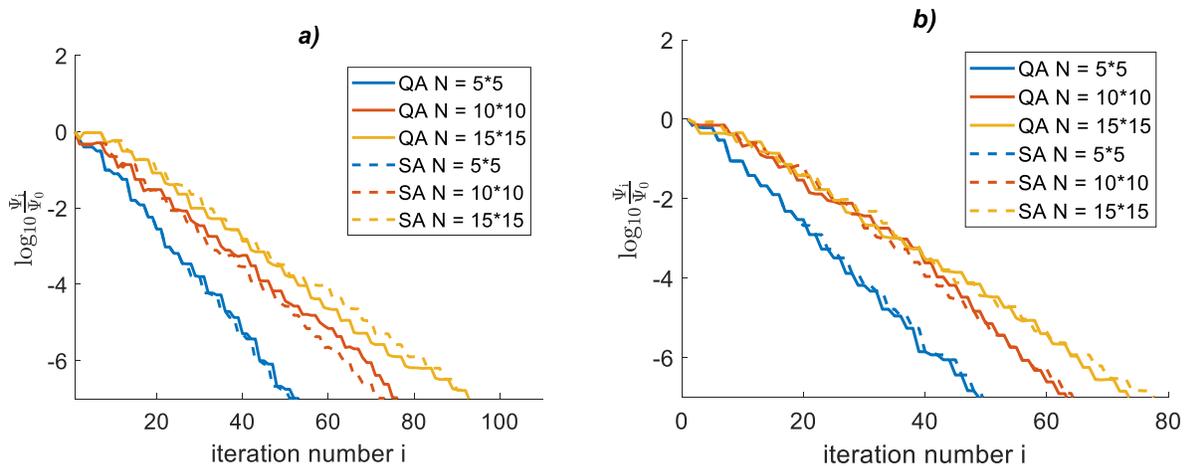

Figure 18 Convergence of the residual for a) Case 1 and b) Case 2

Figure 19 shows the time to target metrics for both simulated and quantum annealing. The energy distributions for quantum annealing and simulated annealing are obtained using the D-Wave 2000Q with default settings and the D-Wave Ocean Library with varying numbers of simulated annealing sweeps, respectively, using 1000 anneals each. 10th percentile TTT metrics are calculated since 10 anneals were used per iteration. The final metrics are calculated by averaging the TTT metrics for 40 random Ising Hamiltonians encountered for 20 randomly selected iteration numbers for both steps of hyperoctant search, except for the smallest problem size for Case 1 where all the Ising Hamiltonians are considered for all the iterations since the total number of iterations is less than 20. Although both quantum and simulated annealing converge in roughly the same order of iterations, a clear advantage in total annealing wall-clock time is seen for quantum annealing.

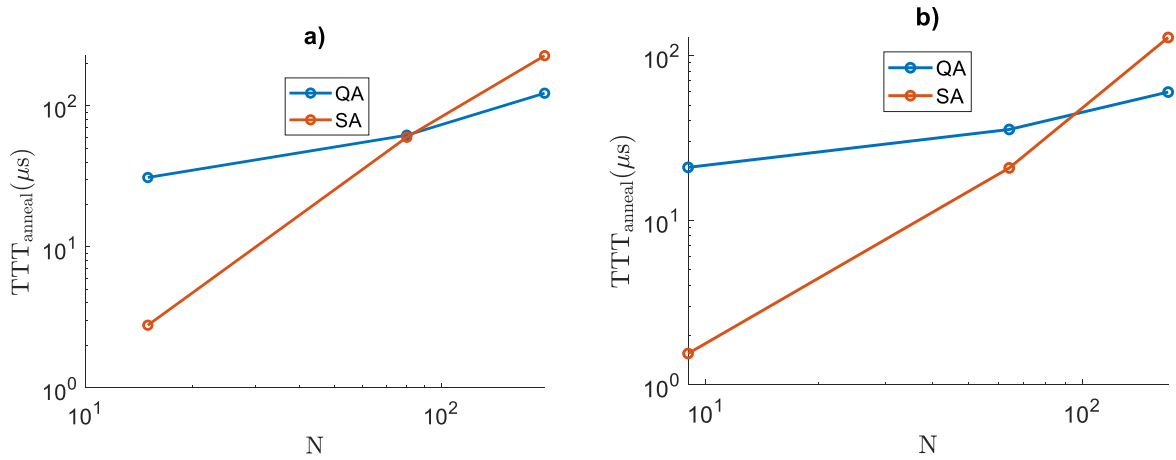

Figure 19 Time to Target metrics for 1D heat transfer in a 2D Domain for a) Case 1 and b) Case 2

Figure 20 shows a comparison of the estimated total wall-clock annealing time for simulated and quantum annealing to reach a normalized residual of $10^{-5}$.

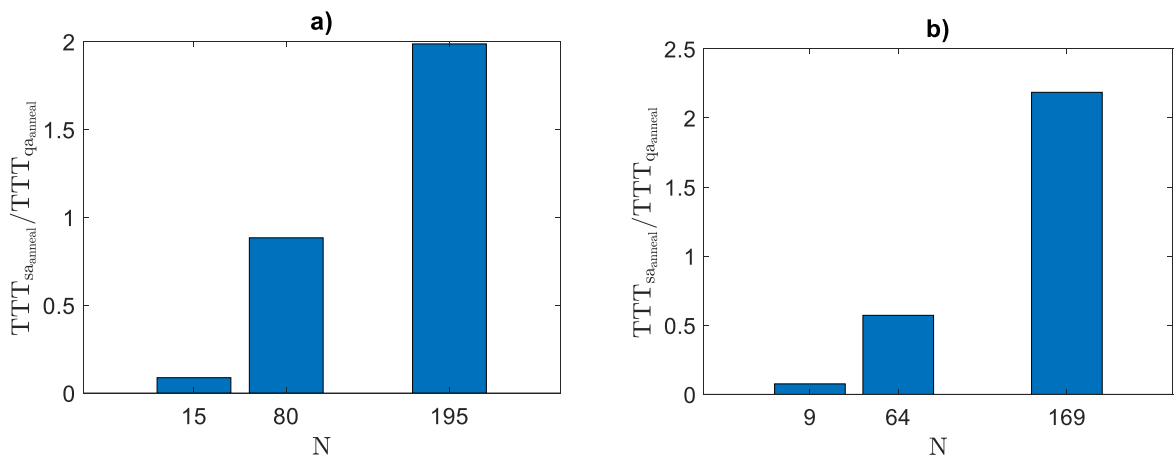

Figure 20 Ratio of total annealing times for simulated and quantum annealing to solve to a precision of $\left|\frac{\Psi_i}{\Psi_0}\right| \leq 10^{-5}$ for a) Case 1 and b) Case 2

# 6. Concluding remarks

Feqa is introduced as a technique for the numerical solution of problems arising from finite element discretizations. Convergence is shown for a variety of problems, and the potential for scaling the methodology is demonstrated. The small number of anneals needed for each iteration allows quantum annealing to be a viable method to solve relatively large problems. A clear advantage is demonstrated over simulated annealing for larger problems using the total anneal time. It is observed that the number of iterations grows slowly with N for some problems. However, the qubit counts and connectivity graphs of currently available annealer hardware restrict the sizes of the solvable problems.

Higher qubit counts and denser graphs will allow more complex meshes and problems to be solved on quantum annealers as hardware improves. Future annealer hardware is expected to implement a non-stoquastic Hamiltonian that will improve performance (Ozfidan et al., 2020; Preskill, 2018). The possibility of qudits in future hardware would allow simpler embeddings without the need for nested iterations to perform $3^N$ and $4^N$ search while resolving the issue of skewed probabilities for $3^N$ search using nested iterations.

A limitation of our approach is that the matrix $\boldsymbol{D}_i$ is restricted to an identity matrix with some non-zero entries eliminated to enforce boundary conditions. Removing this restriction to allow $\boldsymbol{D}_i$ to be a general matrix that still enforces the boundary conditions will allow the search neighborhood to be rotated and stretched. This opens the door to possibilities of using the known properties of a matrix to optimize the search directions or an adaptive algorithm to use information from previous iterations for faster convergence. The presented methodology may also be extended to nonlinear problems through linearization.

The performance of Feqa may also be boosted by using classical techniques for better initial guesses. Using a low-cost classical method to obtain a low precision initial guess and using Feqa to improve upon that solution may be a viable option. Feqa may also be used to improve upon fully converged solutions on a classical computer by directly searching for better solutions near the classical solution that classical solvers cannot find due to floating point discretization errors.

# 7. Acknowledgements

The authors would like to acknowledge the support of this work by NIBIB R01EB0005807, R01EB25241, and NCI R01CA197491 grants.